\documentclass[aps,showpacs,twocolumn,superscriptaddress]{revtex4-2}

\usepackage{amsmath}
\usepackage{amssymb}
\usepackage{graphicx}
\usepackage{pdfsync}
\usepackage{lipsum}
\usepackage{bbm}

\usepackage{siunitx}
\usepackage{epstopdf}
\usepackage{times,txfonts}
\usepackage{easybmat}
\usepackage{mathtools}
\usepackage{upgreek}
\usepackage{appendix}
\newcommand{\ket}[1]{|#1\rangle}

\usepackage[bookmarks=false]{hyperref}
\hypersetup{colorlinks=true,citecolor=blue,linkcolor=blue,urlcolor=blue,pdfstartview=FitH,bookmarksopen=true}
\usepackage{color,soul}
\usepackage{orcidlink}



\begin{document}

\title{Quantum Lock-In Detection via Successive Adiabatic Evolution}
\author{Kangze Li\orcidlink{0000-0001-5258-1167}}
\affiliation{College of Physics and Optoelectronic Engineering and Shanxi Key Laboratory of Precision Measurement Physics, Taiyuan University of Technology, Taiyuan 030024, China}
\affiliation{School of Mathematical Sciences and Centre for the Mathematics and Theoretical Physics of Quantum Non-Equilibrium Systems, University of Nottingham, University Park, Nottingham NG7 2RD, United Kingdom}

\author{Siqi Chen}
\affiliation{School of Physics, State Key Laboratory of Crystal Materials, Shandong University, Jinan, 250100, China}

\author{Hao Zhang}
\affiliation{College of Physics and Optoelectronic Engineering and Shanxi Key Laboratory of Precision Measurement Physics, Taiyuan University of Technology, Taiyuan 030024, China}

\author{Jiazhao Tian}
\affiliation{College of Physics and Optoelectronic Engineering and Shanxi Key Laboratory of Precision Measurement Physics, Taiyuan University of Technology, Taiyuan 030024, China}

\author{Liantuan Xiao\orcidlink{0000-0002-0104-8634}}
\email{xlt@sxu.edu.cn}

\affiliation{College of Physics and Optoelectronic Engineering and Shanxi Key Laboratory of Precision Measurement Physics, Taiyuan University of Technology, Taiyuan 030024, China}

\affiliation{Key Laboratory of Advanced Transducers and Intelligent Control System, Taiyuan University of Technology and Ministry of Education, Taiyuan 030024, China}

\affiliation{State Key Laboratory of Quantum Optics Technologies and Devices, Shanxi University, Taiyuan, 030006, China}

\author{Gerardo Adesso\orcidlink{0000-0001-7136-3755}}
\email{gerardo.adesso@nottingham.ac.uk}
\affiliation{School of Mathematical Sciences and Centre for the Mathematics and Theoretical Physics of Quantum Non-Equilibrium Systems, University of Nottingham, University Park, Nottingham NG7 2RD, United Kingdom}
\date{\today}

\begin{abstract}
In recent years, quantum lock-in detection has emerged as a promising technique to accurately detect weak signals submerged in background noise. However, the signal-to-noise ratio of existing protocols is severely limited by spectral leakage resulting from control operations implemented in pulse form. Here, we propose a general protocol for realizing quantum lock-in detection by employing successive quantum adiabatic evolution. In our protocol, the signal modulation is achieved by adiabatically controlling the time evolution of the quantum probe, which enables the implementation of sinusoidal modulation functions. The realization of sinusoidal-wave modulation substantially mitigates the problem of spectral leakage and facilitates the extraction of the complete characteristics of the target signals. We present a practical implementation scheme of adiabatic quantum lock-in detection based on nitrogen-vacancy centers in diamond, and demonstrate that the proposed protocol possesses strong resilience against experimental imperfections. Our results establish adiabatic quantum lock-in detection as a robust and experimentally accessible approach to detection of weak alternating signals in noisy environments, thus promoting the advance of real-world quantum sensing technologies.
\end{abstract}
\maketitle

\section{Introduction}

The ability to accurately detect noise-submerged signals holds substantial significance in fundamental scientific research and technological advancement \cite{Walsworth21,PRL2007,NRP2024,NP2011,NRM2024,IEEE1986}. In general, when the target signal and the background noise share the same physical essence and couple to the probe via the same physical mechanism, it becomes particularly challenging to enhance the detection sensitivity while simultaneously minimizing the influence of noise. In order to extract the signal from noise with a high signal-to-noise ratio (SNR), the classical {\it lock-in detection} technique has been proposed and widely applied \cite{Cosens1934,Michels1938,Michels1941,Barrios2017,Kishore2020,Zhang2024,Stimpson2019,Gervasoni2017}. Typically, classical lock-in amplifiers isolate the target signal from noise by multiplying the noise-containing signal with reference signals of a specific frequency and subsequently eliminating unwanted frequency components through low-pass filtering. The lock-in detection technique enables simultaneous extraction of the amplitude and initial phase of the target signal.

Although classical lock-in detection technology has matured, the demand for improving detection sensitivity and resolution is ever-growing. For instance, novel techniques such as microscale magnetic resonance spectroscopy \cite{DJF2024,dxb2023} and magnetometry-based characterization of condensed matter materials \cite{pr2016,Casola2018} require precise measurement of electromagnetic signals at the micrometer and even nanometer scales. However, the detection of electromagnetic signals based on classical electronic technology generally relies on macroscopic antennas or inductive structures, which fundamentally restricts the achievable sensitivity and spatial resolution, ultimately exposing the inadequacy of classical techniques for present-day applications.

Quantum technologies may offer a way forward. In recent years, quantum mechanical systems have been used as highly sensitive sensors for various physical quantities due to their high sensitivity to external perturbations and their potential for achieving unprecedented spatial resolution \cite{RMP2017,PR2025,Lvovsky2026,Liao2025,Chaudhry2025,Genov20,Schmitt17,Glenn18,Zeng2024,Cappellaro22,Boss2017,DU2024,ZHANGLJ2024,RB20,TI21,Louzon2025,Mihailescu25}. To realize quantum-enhanced weak-signal detection, some efforts have been made to develop the quantum counterpart of classical lock-in amplifiers \cite{Nature2011,Zhuang2024,CP2024,PRX2021,PRA2021,WEI2025,Shaniv2017}. Several quantum lock-in detection protocols have been applied in magnetic field sensing \cite{Nature2011}, weak-force detection \cite{Shaniv2017}, precise frequency measurement \cite{PRX2021}, and vector light shift detection \cite{PRA2021}. The crux to realizing quantum lock-in detection is to mimic the signal multiplication and filtering operations within the quantum regime. In existing quantum lock-in detection protocols, these processes are usually achieved by utilizing non-commutative operations and time-evolution, respectively. Specifically, the signal multiplication process is realized by implementing periodic sequences of $\pi$ pulses to the quantum probe, which do not commute with the signal Hamiltonian. This means that in these existing quantum lock-in detection protocols, the modulation functions take the form of bipolar square waves rather than the sinusoidal waves used in classical lock-in amplifiers, making it difficult to extract the complete characteristics of the target signals when the initial phases are unknown. Moreover, bipolar square-wave modulation suffers from the significant challenge of spectral leakage, wherein unwanted signal components deviating from the reference frequency can also contribute to the detector’s response and may seriously impair the detection process \cite{PRX2015,NC2017}. Although continuous modulation based on single-axis control has been proposed to mitigate spectral leakage \cite{NC2017}, such a single-axis control method is sensitive to control errors, thereby degrading the performance of the detection process. Therefore, the design and implementation of quantum lock-in detection protocols able to exhibit the advantages of their classical counterparts remain a crucial open issue in the field of quantum sensing.

In this paper, we address this crucial gap by proposing a general protocol for the realization of quantum lock-in detection based on quantum adiabatic evolution \cite{born,Mess,berry}. In contrast to the existing protocols, the proposed adiabatic lock-in detection protocol achieves the signal-reference multiplication process by adiabatically controlling the time-evolution of the quantum probe. Through the implementation of elaborate successive adiabatic evolution processes, the signal multiplication process with two orthogonal sinusoidal modulation functions can be achieved, which is pivotal for enhancing the performance of quantum lock-in detection. The adoption of sinusoidal modulation functions avoids the spectral leakage resulting from control operations implemented in pulse form, hence enabling the adiabatic quantum lock-in detection to selectively filter out unwanted frequency components more effectively. Moreover, our proposed adiabatic lock-in detection protocol is capable of directly extracting the complete characteristics of the target signals, including amplitude, frequency, and initial phase, without the need for complicated post-processing procedures. In addition, our adiabatic quantum lock-in detection protocol is also more viable in experimental settings because adiabatic evolution is inherently robust against parameter variations and timing errors. The robustness of adiabatic quantum lock-in detection can be further enhanced by integrating it with the composite pulse technique \cite{Levitt1986,Tycko1984,Torosov2011,Genov2020,Paspalakis19}. Thus, the adiabatic quantum lock-in detection protocol can appreciably extend the coherence time of quantum probes and ultimately increase the detection sensitivity.

To demonstrate the feasibility and advantages of the proposed protocol, we consider its application to the detection of nuclear magnetic resonance (NMR) signals from nuclear spin ensembles in micron-sized samples, where the NMR signals can be effectively regarded as classical AC magnetic fields \cite{Laraoui11, Aslam17, Munuera17}. For this application scenario, we present a detailed implementation scheme utilizing nitrogen-vacancy (NV) centers in diamond. The excellent spectral selectivity of the adiabatic quantum lock-in detection protocol offers a promising approach to addressing the long-standing challenge of spurious resonance in NMR spectroscopy \cite{Zeng2024,PRX2015}. Furthermore, we quantitatively characterize the advantages of the proposed protocol over conventional pulse-based protocols in terms of frequency selectivity and noise resistance via both analytical derivation and numerical simulations. Our analysis demonstrates that the proposed protocol significantly outperforms conventional pulse-based protocols, providing a promising pathway for the practical implementation and application of lock-in detection in microscale NMR spectroscopy and other quantum sensing technologies.

The paper is organized as follows. In Section \ref{GP}, we introduce our general protocol for quantum lock-in detection based on successive adiabatic evolution. In Section \ref{IS}, we discuss microscale NMR spectroscopy as a representative application and present a practical implementation using NV centers in diamond. In Section \ref{AD}, we analyze the frequency selectivity, noise robustness, control errors, decoherence, and precision of the protocol. In Section \ref{CON}, we present a brief conclusion.

\section{GENERAL PROTOCOL}\label{GP}

The goal of the lock-in detection technique is to extract a target signal from extreme environmental noise. Specifically, a target signal with a known carrier frequency $\omega_{c}$ takes the form
\begin{align}
S(t)=S_{0}\cos(\omega_{c}t+\varphi),\label{eq_signal}
\end{align}
 where the amplitude $S_{0}$ and the initial phase $\varphi$ are parameters to be measured. The target signal is submerged in noise $N(t)$, so the actual signal sensed by the detector is
 \begin{align}
 M(t)=S(t)+N(t).\label{eq_signalwithnoise}
 \end{align}

Classical lock-in amplifiers achieve their function through a mix-down process. The actual signal $M(t)$ is multiplied by the reference signals $\sin(\omega t)$ and $\cos(\omega t)$, and then integrated over a period $T$. Consequently, the lock-in signals can be expressed as
\begin{align}
\begin{split}
I_{\text{sig}}=\frac{1}{T}\int^{T}_{0}M(t)\cos(\omega t) dt,\\
Q_{\text{sig}}=\frac{1}{T}\int^{T}_{0}M(t)\sin(\omega t) dt,
\end{split}
     \label{eq0}
\end{align}
in which $I_{\text{sig}}$ is the in-phase component, and $Q_{\text{sig}}$ is the quadrature component.

In the mix-down process, the noise with frequencies significantly different from the reference frequency $\omega$ is averaged out. Under the condition that the reference frequency equals the carrier frequency ($\omega = \omega_c$), information about the target signal can be extracted by measuring the lock-in signals in Eq.~(\ref{eq0}). Specifically, the amplitude of the target signal can be obtained by $S_{0}=2(I_{\text{sig}}^{2}+Q_{\text{sig}}^{2})^{1/2}$, and the initial phase can be obtained by $\varphi=\arctan(-Q_{\text{sig}}/I_{\text{sig}})$, where the quadrant ambiguity can be resolved by using the signs of $I_{\text{sig}}$ and $Q_{\text{sig}}$.

Generally, the classical mix-down process is implemented by using certain nonlinear devices that are able to multiply two input signals. In the quantum regime, the nonlinear dynamics of the wave function cannot be introduced directly due to the linearity of the Schr\"{o}dinger equation. Therefore, in the previously proposed quantum lock-in detection protocols, the quantum mix-down process is realized using non-commuting dynamical manipulations implemented in the form of pulse sequences \cite{Nature2011,PRX2021,Shaniv2017}. However, such pulse manipulations are significantly affected by spectral leakage. To solve this issue, we propose a novel quantum lock-in detection mechanism that supports sinusoidal modulation and thereby avoids spectral leakage. In the following sections, we illustrate the building blocks of our quantum lock-in detection protocol in detail.

\subsection{Adiabatic evolution of the quantum probe}
To realize quantum lock-in detection, we consider a driven two-level quantum probe subject to the actual signal $M(t)$, which can generally be described by the Hamiltonian under the rotating-wave approximation
\begin{align}
\begin{split}
H(t)&=H_{\text{ctrl}}(t)+H_{\text{sig}}(t)\\
 &=\left[\frac{\Omega_{R}(t)}{2}(\cos\phi\sigma_{x}+\sin\phi\sigma_{y})+\frac{\Delta(t)}{2}\sigma_{z}\right]+M(t)\sigma_{z},
\end{split}
     \label{eq1}
\end{align}
in which $\sigma_{x}$, $\sigma_{y}$, and $\sigma_{z}$ are the standard Pauli operators. The control Hamiltonian $H_{\text{ctrl}}(t)$ represents a control field with Rabi frequency $\Omega_{R}(t)$, detuning $\Delta(t)$, and phase $\phi$.  The signal Hamiltonian $H_{\text{sig}}(t)=M(t)\sigma_{z}$ represents the coupling between the quantum probe and the actual signal $M(t)$. Here, we consider classical signals, such as magnetic fields, that interact with the quantum probe via the Zeeman effect.

We assume that the actual signal $M(t)$ is weak compared to the control field, and thus $H_{\text{sig}}(t)$ can be regarded as a perturbation term. Our purpose is to achieve quantum lock-in detection by using the adiabatic evolution driven by $H_{\text{ctrl}}(t)$. We first consider the adiabatic evolution driven by the unperturbed term $H_{\text{ctrl}}(t)$ in Hamiltonian (\ref{eq1}). The instantaneous eigenstates of $H_{\text{ctrl}}(t)$ take the form of
\begin{align}
\begin{split}
&\ket{E_{1}(t)}=\cos\frac{\theta(t)}{2}~\ket{0}+\sin\frac{\theta(t)}{2}e^{i\phi}~\ket{1},\\
&\ket{E_{2}(t)}=-\sin\frac{\theta(t)}{2}e^{-i\phi}~\ket{0}+\cos\frac{\theta(t)}{2}~\ket{1},
\end{split}
     \label{eq2}
\end{align}
in which the mixing angle $\theta(t)$ satisfies $\cos\theta(t)=\Delta(t)/\Omega(t)$ with $\Omega(t)=[\Delta^{2}(t)+\Omega_{R}^{2}(t)]^{\frac{1}{2}}$; $\ket{E_{1}(t)}$ and $\ket{E_{2}(t)}$ correspond to eigenvalues $E_{1}(t)=\Omega(t)/2$ and $E_{2}(t)=-\Omega(t)/2$, respectively. According to the adiabatic theorem, if $H_{\text{ctrl}}(t)$ varies slowly enough, the system initially in the eigenstate $\ket{\psi_{n}(t_{0})}=\ket{E_{n}(t_{0})}$ will evolve transitionlessly along the state
\begin{align}
\ket{\psi_{n}(t)}=e^{-i\int^{t}_{t_{0}}E_{n}(t^{\prime})dt^{\prime}-\int^{t}_{t_{0}}\langle E_{n}(t^{\prime})|\partial_{t^{\prime}}E_{n}(t^{\prime})\rangle dt^{\prime}}\ket{E_{n}(t)}.
     \label{eq3}
\end{align}
In Eq.~(\ref{eq3}) the effective vector potential $\langle E_{n}(t)|\partial_{t}E_{n}(t)\rangle$ that generates the geometric phase is equal to zero for the eigenstates in Eq.~(\ref{eq2}). This means that the eigenstate $\ket{E_{n}(t)}$ only accumulates dynamical phase $D_{n}(t,t_{0})=-\int^{t}_{t_{0}}E_{n}(t^{\prime})dt^{\prime}$ during the adiabatic evolution. Therefore, in the rotating frame defined by the time-dependent basis $\{e^{iD_{1}}\ket{E_{1}(t)},e^{iD_{2}}\ket{E_{2}(t)}\}$ the propagator takes the form of unit operator
\begin{align}
\begin{split}
U_{\text{rot}}(t,t_{0})=\left(
       \begin{array}{ccc}
         1 &0\\
         0 &1\\
       \end{array}
     \right).
\end{split}
     \label{eq4}
\end{align}
In the basis $\{\ket{0},\ket{1}\}$, the propagator of the adiabatic process can be obtained by
\begin{align}
\begin{split}
U(t,t_{0})=V(t,t_{0})U_{\text{rot}}V^{\dag}(t_{0},t_{0}),
\end{split}
     \label{eq5}
\end{align}
in which
\begin{align}
\begin{split}
V(t,t_{0})=\left(
       \begin{array}{ccc}
         e^{iD_{1}}\cos\frac{\theta(t)}{2} &-e^{i(D_{2}-\phi)}\sin\frac{\theta(t)}{2}\\
         e^{i(D_{1}+\phi)}\sin\frac{\theta(t)}{2} & e^{iD_{2}}\cos\frac{\theta(t)}{2}\\
       \end{array}
     \right)
\end{split}
     \label{eq6}
\end{align}
is the unitary operator associated with the basis change.

Then, we consider the effect of the signal Hamiltonian $H_{\text{sig}}(t)$ in the rotating frame defined by the time-dependent basis $\{e^{iD_{1}}\ket{E_{1}(t)},e^{iD_{2}}\ket{E_{2}(t)}\}$. Due to the presence of $H_{\text{sig}}(t)$, the propagator in the rotating frame is no longer the unit operator and takes the form of
\begin{align}
U_{\text{rot}}(t,t_{0})=\mathcal{T}\exp\left[-i\int^{t}_{t_{0}}H_{\text{rot}}(t^{\prime},t_{0})dt^{\prime}\right]
     \label{eq7}
\end{align}
in which $H_{\text{rot}}(t,t_{0})$ is the Hamiltonian in the rotating frame with
\begin{align}
\begin{split}
H_{\text{rot}}(t,t_{0})&=M(t)V^{\dag}(t,t_{0})\sigma_{z}V(t,t_{0})\\
&=M(t)\left(
       \begin{array}{ccc}
         \cos\theta &-e^{i(D_{2}-D_{1}-\phi)}\sin\theta\\
         -e^{i(D_{1}-D_{2}+\phi)}\sin\theta & -\cos\theta\\
       \end{array}
     \right).
\end{split}
     \label{eq8}
\end{align}

In the following, we shall demonstrate that the quantum mix-down process can be achieved by subjecting the quantum probe to successive adiabatic evolution processes.

\subsection{Measurement of the in-phase component}\label{GPB}
The in-phase component $I_{\text{sig}}$ in Eq.~(\ref{eq0}) can be measured using a sequence consisting of two successive adiabatic evolution processes. The first adiabatic process occurs during the time interval $t\in{[0,\tau]}$, and the second adiabatic process occurs during the time interval $t\in{(\tau,T]}$, where $\tau=T/2$. In the first adiabatic process, the Rabi frequency, detuning, and phase of the control field are set as $\Omega_{R}(t)=A\sin(\omega t)$, $\Delta(t)=A\cos(\omega t)$, and $\phi=\phi_{1}$, which corresponds to the mixing angle $\theta(t)=\omega t$. The parameter $A$ is set to be much larger than the highest frequency $\omega_{\mathrm{max}}$ present in the noise $N(t)$.

By using the first-order Magnus expansion \cite{mg}, the approximate propagator in the rotating frame at the time $\tau=\pi/\omega$ can be expressed as
\begin{align}
\begin{split}
U_{\text{rot}}^{m}(\tau,0)&=\exp\left[-i\int^{\tau}_{0}H_{\text{rot}}(t^{\prime},0)dt^{\prime}\right]\\
&=\left(
       \begin{array}{ccc}
         e^{-i\int^{\tau}_{0}M(t)\cos(\omega t) dt} &0\\
         0 & e^{i\int^{\tau}_{0}M(t)\cos(\omega t) dt}\\
       \end{array}
     \right).
\end{split}
     \label{eq9}
\end{align}
It is worth noting that the nondiagonal terms in Eq.~(\ref{eq8}) do not contribute to the propagator in Eq.~(\ref{eq9}). This is because when $A\gg\omega_{\text{max}}$, these terms oscillate rapidly and their effects are averaged out during the evolution process. By using Eqs.~(\ref{eq5}) and (\ref{eq9}), the propagator of the first adiabatic process in the basis $\{\ket{0},\ket{1}\}$ can be obtained as

\begin{align}
\begin{split}
U(\tau,0)=\left(
       \begin{array}{ccc}
         0 &-e^{i[\frac{A\tau}{2}-\phi_{1}+\alpha(\tau,0)]}\\
         e^{i[-\frac{A\tau}{2}+\phi_{1}-\alpha(\tau,0)]} & 0\\
       \end{array}
     \right),
\end{split}
     \label{eq10}
\end{align}
in which $\alpha(\tau,0)=\int^{\tau}_{0}M(t)\cos(\omega t)dt$.

In the second adiabatic process, the Rabi frequency, detuning, and phase of the control field are set as $\Omega_{R}(t)=A\sin(\omega t-\pi)$, $\Delta(t)=A\cos(\omega t-\pi)$, and $\phi=\phi_{2}$, which corresponds to the mixing angle $\theta(t)=\omega t-\pi$. The approximate propagator in the rotating frame can be obtained as

\begin{align}
\begin{split}
U_{\text{rot}}^{m}(T,\tau)=\left(
       \begin{array}{ccc}
         e^{i\int^{T}_{\tau}M(t)\cos(\omega t)dt} &0\\
         0 & e^{-i\int^{T}_{\tau}M(t)\cos(\omega t)dt}\\
       \end{array}
     \right).
\end{split}
     \label{eq11}
\end{align}
The propagator of the second adiabatic process in the basis $\{\ket{0},\ket{1}\}$ can be obtained as

\begin{align}
\begin{split}
U(T,\tau)=\left(
       \begin{array}{ccc}
         0 &-e^{i[\frac{A\tau}{2}-\phi_{2}-\alpha(T,\tau)]}\\
         e^{i[-\frac{A\tau}{2}+\phi_{2}+\alpha(T,\tau)]} & 0\\
       \end{array}
     \right),
\end{split}
     \label{eq12}
\end{align}
in which $\alpha(T,\tau)=\int^{T}_{\tau}M(t)\cos(\omega t)dt$.

Therefore, the propagator of the two successive adiabatic evolution processes in the basis $\{\ket{0},\ket{1}\}$ takes the form
\begin{align}
\begin{split}
U(T,0)&=U(T,\tau)U(\tau,0)\\
&=\left(\begin{array}{ccc}
         -e^{-i[\phi_{2}-\phi_{1}+\alpha(T,0)]} & 0\\
         0 & -e^{i[\phi_{2}-\phi_{1}+\alpha(T,0)]}\\
       \end{array}
     \right),
\end{split}
     \label{eq13}
\end{align}
in which
\begin{align}
\alpha(T,0)=\int^{T}_{0}M(t)\cos(\omega t)dt.
     \label{eq14}
\end{align}
This indicates that if the quantum probe is initialized into the superposition state $\ket{\psi(0)}=\frac{1}{\sqrt{2}}(\ket{0}+\ket{1})$ and undergoes the two successive adiabatic evolutions, the superposition state will pick up a relative phase of $2[\phi_{2}-\phi_{1}+\alpha(T,0)]$. The phases $\phi_{1}$ and $\phi_{2}$ are known, and $\alpha(T,0)$ in Eq.~(\ref{eq14}) differs from $I_{\text{sig}}$ in Eq.~(\ref{eq0}) only by a constant factor. Thus, the in-phase component $I_{\text{sig}}$ can be measured by detecting the relative phase accumulated during the two successive adiabatic evolutions.

\subsection{Measurement of the quadrature component}\label{Q}

The quadrature component $Q_{\text{sig}}$ in Eq.~(\ref{eq0}) can also be measured using a sequence consisting of three successive adiabatic evolution processes. The first adiabatic process occurs during the time interval $t\in{[0,\tau/2]}$, the second adiabatic process occurs during the time interval $t\in{(\tau/2,3\tau/2)}$, and the third adiabatic process occurs during the time interval $t\in{[3\tau/2,T]}$, where $\tau=T/2$.

In the first adiabatic evolution process, the Rabi frequency, detuning, and phase of the control field are set as $\Omega_{R}(t)=A\sin(\omega t+\pi/2)$, $\Delta(t)=A\cos(\omega t+\pi/2)$, and $\phi=\phi_{1}$, which corresponds to the mixing angle $\theta(t)=\omega t+\pi/2$. By using the first-order Magnus expansion \cite{mg}, the approximate propagator in the rotating frame at the time $\tau/2=\pi/2\omega$ can be obtained as
\begin{align}
\begin{split}
U_{\text{rot}}^{m}\left(\frac{\tau}{2},0\right)=\left(
       \begin{array}{ccc}
         e^{i\int^{\tau/2}_{0}M(t)\sin(\omega t)dt} &0\\
         0 & e^{-i\int^{\tau/2}_{0}M(t)\sin(\omega t)dt}\\
       \end{array}
     \right).
\end{split}
     \label{eq15}
\end{align}
By using Eqs.~(\ref{eq5}) and (\ref{eq15}), the propagator of the first adiabatic evolution process in the basis $\{\ket{0},\ket{1}\}$ can be obtained as

\begin{align}
\begin{split}
U\left(\frac{\tau}{2},0\right)=\left(
       \begin{array}{ccc}
         \frac{\sqrt{2}}{2}e^{i[\frac{A\tau}{4}-\beta(\tau/2,0)]} &-\frac{\sqrt{2}}{2}e^{i[\frac{A\tau}{4}-\phi_{1}-\beta(\tau/2,0)]}\\
         \frac{\sqrt{2}}{2}e^{i[-\frac{A\tau}{4}+\phi_{1}+\beta(\tau/2,0)]} &\frac{\sqrt{2}}{2}e^{i[-\frac{A\tau}{4}+\beta(\tau/2,0)]}\\
       \end{array}
     \right),
\end{split}
     \label{eq16}
\end{align}
where $\beta(\tau/2,0)=\int^{\tau/2}_{0}M(t)\sin(\omega t)dt$.

In the second adiabatic evolution process, the Rabi frequency, detuning, and phase of the control field are set as $\Omega_{R}(t)=A\sin(\omega t-\pi/2)$, $\Delta(t)=A\cos(\omega t-\pi/2)$, and $\phi=\phi_{2}$, which corresponds to the mixing angle $\theta(t)=\omega t-\pi/2$. The approximate propagator in the rotating frame can be obtained as
\begin{align}
\begin{split}
U_{\text{rot}}^{m}\left(\frac{3\tau}{2},\frac{\tau}{2}\right)=\left(
       \begin{array}{ccc}
       e^{-i\int^{3\tau/2}_{\tau/2}M(t)\sin(\omega t)dt}& 0\\
        0 & e^{i\int^{3\tau/2}_{\tau/2}M(t)\sin(\omega t)dt}\\
       \end{array}
     \right).
\end{split}
     \label{eq17}
\end{align}
The propagator of the second adiabatic evolution in the basis $\{\ket{0},\ket{1}\}$ can be obtained as
\begin{align}
\begin{split}
U\left(\frac{3\tau}{2},\frac{\tau}{2}\right)=\left(
       \begin{array}{ccc}
         0 & -e^{i[\frac{A\tau}{2}-\phi_{2}+\beta(3\tau/2,\tau/2)]}\\
        e^{i[-\frac{A\tau}{2}+\phi_{2}-\beta(3\tau/2,\tau/2)]} & 0\\
       \end{array}
     \right),
\end{split}
     \label{eq18}
\end{align}
where $\beta(3\tau/2,\tau/2)=\int^{3\tau/2}_{\tau/2}M(t)\sin(\omega t)dt$.

In the third adiabatic evolution process, the Rabi frequency, detuning, and phase of the control field are set as $\Omega_{R}(t)=A\sin(\omega t+\pi/2)$, $\Delta(t)=A\cos(\omega t+\pi/2)$, and $\phi=\phi_{3}$, which corresponds to the mixing angle $\theta(t)=\omega t+\pi/2$. The approximate propagator in the rotating frame can be obtained as
\begin{align}
\begin{split}
U_{\text{rot}}^{m}\left(T,\frac{3\tau}{2}\right)=\left(
       \begin{array}{ccc}
       e^{i\int^{T}_{3\tau/2}M(t)\sin(\omega t)dt}& 0\\
        0 & e^{-i\int^{T}_{3\tau/2}M(t)\sin(\omega t)dt}\\
       \end{array}
     \right).
\end{split}
     \label{eq19}
\end{align}
The propagator of the third adiabatic evolution in the basis $\{\ket{0},\ket{1}\}$ can be obtained as
\begin{align}
\begin{split}
U\left(T,\frac{3\tau}{2}\right)=\left(
       \begin{array}{ccc}
         \frac{\sqrt{2}}{2}e^{i[-\frac{A\tau}{4}+\beta(T,3\tau/2)]} &-\frac{\sqrt{2}}{2}e^{i[\frac{A\tau}{4}-\phi_{3}-\beta(T,3\tau/2)]}\\
         \frac{\sqrt{2}}{2}e^{i[-\frac{A\tau}{4}+\phi_{3}+\beta(T,3\tau/2)]} &\frac{\sqrt{2}}{2}e^{i[\frac{A\tau}{4}-\beta(T,3\tau/2)]}\\
       \end{array}
     \right),
\end{split}
     \label{eq20}
\end{align}
where $\beta(T,3\tau/2)=\int^{T}_{3\tau/2}M(t)\sin(\omega t)dt$.

Therefore, the propagator of the three successive adiabatic evolution processes in the basis $\{\ket{0},\ket{1}\}$ takes the form of

\begin{widetext}
\begin{equation}
\begin{split}
U(T,0)=U\left(T,\frac{3\tau}{2}\right)U\left(\frac{3\tau}{2},\frac{\tau}{2}\right)U\left(\frac{\tau}{2},0\right)=\left(\begin{array}{ccc}
         -\frac{1}{2}e^{i[\phi_{2}-\phi_{3}-\beta(T,0)]}-\frac{1}{2}e^{i[\phi_{1}-\phi_{2}+\beta(T,0)]} & \frac{1}{2}e^{i[\phi_{2}-\phi_{1}-\phi_{3}-\beta(T,0)]}-\frac{1}{2}e^{i[-\phi_{2}+\beta(T,0)]}\\
          \frac{1}{2}e^{i[\phi_{2}-\beta(T,0)]}-\frac{1}{2}e^{i[\phi_{1}+\phi_{3}-\phi_{2}+\beta(T,0)]} & -\frac{1}{2}e^{i[\phi_{2}-\phi_{1}-\beta(T,0)]}-\frac{1}{2}e^{i[\phi_{3}-\phi_{2}+\beta(T,0)]}\\
       \end{array}\right),
\end{split}
     \label{eq21}
\end{equation}
\end{widetext}
in which
\begin{equation}
\beta(T,0)=\int^{T}_{0}M(t)\sin(\omega t) dt.
     \label{eq22}
\end{equation}
The phases $\phi_{1}$, $\phi_{2}$, and $\phi_{3}$ are known, and $\beta(T,0)$ in Eq.~(\ref{eq22}) differs from $Q_{\text{sig}}$ in Eq.~(\ref{eq0}) only by a constant factor. Thus, the quadrature component $Q_{\text{sig}}$ can be obtained by measuring the final state of the quantum probe after the three successive adiabatic evolutions. For instance, under the condition that $\phi_{1}=\phi_{3}=\pi/2$, $\phi_{2}=0$ the propagator in Eq.~(\ref{eq21}) reduces to
\begin{equation}
\begin{split}
U(T,0)=\left(\begin{array}{ccc}
         \sin\beta(T,0) & -\cos\beta(T,0)\\
         \cos\beta(T,0) & \sin\beta(T,0)\\
       \end{array}\right),
\end{split}
     \label{eq23}
\end{equation}
If the quantum probe is initialized into the state $\ket{\psi(0)}=\ket{0}$ and undergoes the three successive adiabatic evolutions, the final state will be $\ket{\psi(T)}=\sin\beta(T,0)\ket{0}+\cos\beta(T,0)\ket{1}$.
This indicates that the quadrature component $Q_{\text{sig}}$ can be obtained by measuring the parameter $\beta(T,0)$ after the three successive adiabatic evolutions.

\subsection{Measurement of the target signal with unknown frequency}

In general, it is necessary to know the carrier frequency $\omega_c$ of the target signal for the implementation of lock-in detection, as it determines the frequency $\omega$ of the reference signals. However, in some relevant sensing scenarios, the carrier frequency may be unknown. To address this, we present a method for determining the carrier frequency, enabling the general application of our protocol even in cases where the carrier frequency is unknown. This method is also based on the controlled adiabatic evolution of the quantum probe.

We still consider that the target signal takes the form $S(t)=S_{0}\cos(\omega_{c} t+\varphi)$ as in Eq.~(\ref{eq_signal}), but the carrier frequency $\omega_{c}$, amplitude $S_{0}$, and initial phase $\varphi$ are all unknown. In order to determine the carrier frequency $\omega_{c}$, we consider the adiabatic evolution sequence used to measure the in-phase component in the previous section. If a driven two-level quantum probe undergoes the adiabatic evolution sequence while being subjected to a target signal with an unknown frequency, the phase in Eq.~(\ref{eq14}) will become
\begin{align}
\alpha=S_{0}\int^{T}_{0}\cos(\omega_{c} t+\varphi)\cos(\omega t)dt.
     \label{eq24}
\end{align}
Here, we assume that the frequency of the noise differs significantly from both the carrier frequency $\omega_{c}$ and the reference frequency $\omega$, so its contribution to $\alpha$ is averaged out and can be ignored. In Eq.~(\ref{eq24}), the reference frequency $\omega$ and its corresponding period $T = 2\pi/\omega$ are known, the carrier frequency $\omega_c$ and the amplitude $S_0$ are unknown but fixed across multiple implementation of the adiabatic evolution sequence, and the initial phase $\varphi$ depends on the start time of each sequence. Therefore, the phase $\alpha$ in Eq.~(\ref{eq24}) can be regarded as a function of the variable $\varphi$ with a period of $2\pi$.

The sequences used to measure the in-phase component can be applied in pairs with an adjustable time interval. By comparing the $\alpha$ values obtained from sequences with different time intervals, the carrier frequency $\omega_{c}$ can be determined. The time interval between the start times of the two sequences is denoted as $T^{\prime}$. Due to the periodicity of $\alpha(\varphi)$, for any $T^{\prime}_{n}$ satisfying $\omega_{c}T^{\prime}_{n}=2n\pi, n\in \mathbbm{N}_{+}$, the two adiabatic evolution sequences yield identical values of $\alpha$ across multiple implementations with arbitrary initial phase. In practice, one can approximately identify $T^{\prime}_{n}$ by stepwise adjusting the time interval $T^{\prime}$ and repeatedly measuring the values of $\alpha$ multiple times at each step. The time interval between two consecutive $T^{\prime}_{n}$ is the period of the target signal, denoted as $T_{c}=T^{\prime}_{n+1}-T^{\prime}_{n}$. Thus, the carrier frequency $\omega_{c}$ can then be obtained from the relation $\omega_{c} = 2\pi / T_{c}$. After obtaining the carrier frequency $\omega_{c}$, the reference frequency can be set to $\omega = \omega_{c}$, and then the quantum lock-in detection protocol proposed in the previous section can be applied to extract the amplitude and phase of the target signal.

A discussion is in order regarding the accuracy and time overhead of the frequency estimation process. Let us assume that the time step used in the process of estimating $T_{c}$ is set to $\Delta T^{\prime}$. Because we only need to identify two consecutive $T^{\prime}_{n}$, the bound on the number of steps is given as
\begin{align}
\left\lceil\frac{T_{c}}{\Delta T^{\prime}}\right\rceil<N_{\text{step}}<\left\lceil\frac{2T_{c}}{\Delta T^{\prime}}\right\rceil.
     \label{eq25}
\end{align}
The absolute quantization error in the process of estimating $T_{c}$ equals the time step $\Delta T^{\prime}$. By using the relation $\omega_{c} = 2\pi / T_{c}$, the relationship between the time quantization error $\Delta T^{\prime}$, the absolute frequency error $\Delta \omega_{c}$, and the relative error $\eta$ can be derived as
\begin{align}
\frac{\Delta T^{\prime}}{T_{c}}=\frac{\Delta \omega_{c}}{\omega_{c}}=\eta
     \label{eq26}
\end{align}
By using Eqs.~(\ref{eq25}) and (\ref{eq26}), the upper bound on the number of steps can be reexpressed as
\begin{align}
N_{\text{step}}<\left\lceil\frac{2}{\eta}\right\rceil.
     \label{eq27}
\end{align}
If the maximum allowable relative error is set to $\eta_0$, the inequality
\begin{align}
\frac{T_{c}}{\Delta T^{\prime}}>\frac{1}{\eta_{0}}
     \label{eq28}
\end{align}
must be satisfied. By using Eqs.~(\ref{eq25}) and (\ref{eq28}), a lower bound on the number of steps can be obtained as
\begin{align}
\left\lceil\frac{1}{\eta_{0}}\right\rceil<N_{\text{step}}.
     \label{eq29}
\end{align}
These results indicate that a more accurate frequency estimation is associated with a greater time overhead.

Up to this point, we have theoretically demonstrated that quantum lock-in detection can be achieved via successive adiabatic evolutions. In the following section, we shall present a scheme for the practical implementation of adiabatic quantum lock-in detection.

\begin{figure*}[htb]
  \includegraphics[scale=0.67]{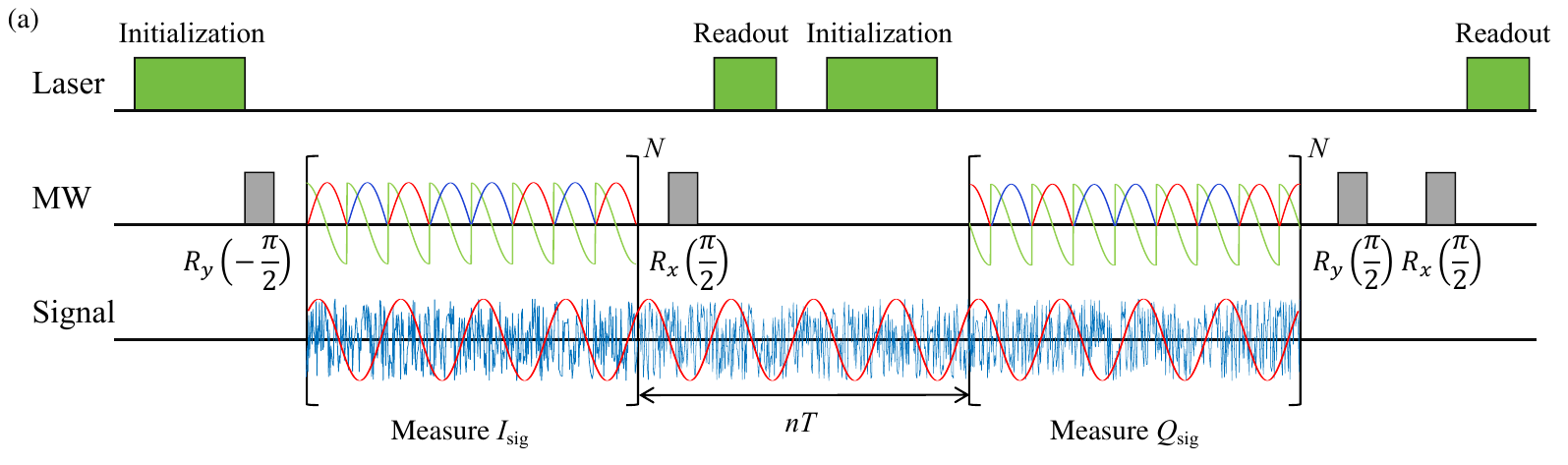}
  \includegraphics[scale=0.8]{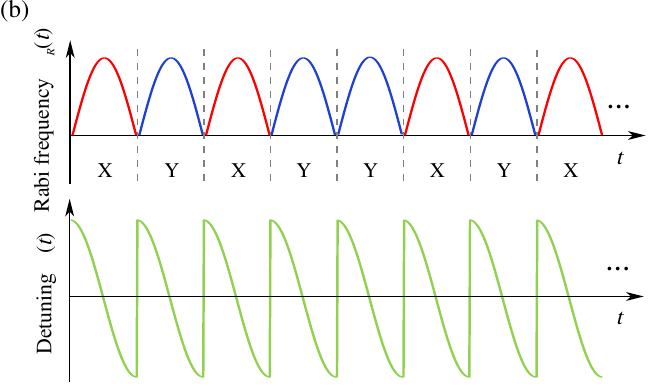}
  \includegraphics[scale=0.8]{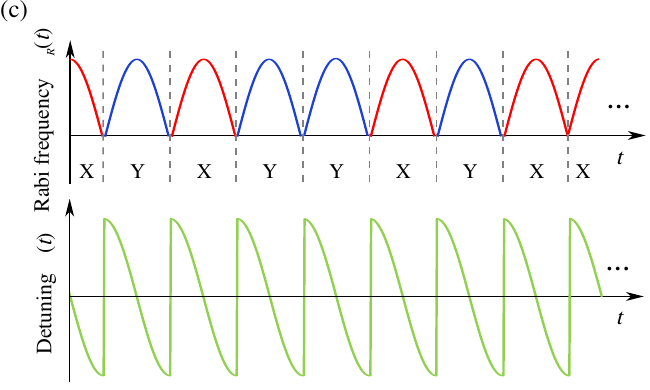}
  \caption{Schematic of adiabatic quantum lock-in detection using NV centers in diamond. (a) Flow diagram of adiabatic quantum lock-in detection. The top row shows the laser pulse sequence used to initialize and read out the electron spin state of NV centers. The middle row shows the driving microwave field used to implement the measurement process and gate operations. The bottom row shows the target signal and noise. (b) Rabi frequency and detuning of the driving microwave field used in the measurement process of the in-phase component $I_{\text{sig}}$. (c) Rabi frequency and detuning of the driving microwave field used in the measurement process of the quadrature component $Q_{\text{sig}}$.}
   \label{Fig1}
\end{figure*}

\section{APPLICATION AND IMPLEMENTATION}\label{IS}

Having introduced the general protocol in Sec.~\ref{GP}, we now illustrate its practical relevance and experimental feasibility in the context of microscale NMR spectroscopy \cite{PRX2015,Laraoui11,Aslam17,Munuera17}, together with a concrete implementation based on NV centers in diamond \cite{Doherty13}.

\subsection{Application in microscale NMR spectroscopy}

NMR spectroscopy is one of the most widely employed analytical techniques across a broad range of research fields including biology, chemistry, medicine, and materials science. Recently, advances in quantum sensing based on NV centers have extended NMR spectroscopy to the micro- and nanoscale. However, spurious resonance resulting from the non-ideal harmonic response of the detection protocol can result in ambiguous signal identification \cite{Zeng2024,PRX2015}. The high spectral selectivity of the proposed adiabatic quantum lock-in protocol is therefore particularly relevant to applications in NMR spectroscopy.

We take the electron spin associated with a single negatively charged NV center in diamond as the quantum probe, with the host $^{15}\mathrm{N}$ nuclear spin polarized \cite{ref10}. The NV center has a spin-triplet ground state with sublevels $\ket{m_{s}=0}$ and $\ket{m_{s}=\pm1}$. Its Hamiltonian is $H_{NV}=DS_{z}^{2}+\gamma_{e}B_{0}S_{z}$, where $D=2\pi\times2.87~\text{GHz}$ is the zero-field splitting and $\gamma_{e}=2\pi\times28~\text{GHz/T}$ is the electron gyromagnetic ratio. A static magnetic field $B_{0}\approx500~\text{G}$ applied along the NV axis lifts the degeneracy of the $\ket{m_{s}=\pm1}$ states by a Zeeman splitting $2\gamma_{e}B_{0}$.

We consider a micron-sized sample containing numerous target nuclear spins weakly coupled to the NV-center electron spin through dipolar interactions \cite{Aslam17,Munuera17}. In this regime, a semiclassical description of the nuclear-spin ensemble is applicable \cite{Laraoui11}, in which the back-action of the nuclear spin ensemble is neglected. As the target nuclear spins precess around the static field $B_{0}$ at frequency $\omega_{c}$, their summed dipolar field can be represented as an AC magnetic signal $b_{s}(t)=b_{0}\cos(\omega_{c}t+\varphi)$. Detecting this field provides information on the chemical composition, physical structure, or dynamics of the sample.

For sensing, we encode the two-level probe in $\ket{m_{s}=-1}\equiv\ket{0}$ and $\ket{m_{s}=0}\equiv\ket{1}$. In this subspace, the Hamiltonian reduces to $H_{NV}=\omega_{0}\sigma_{z}/2$, with $\omega_{0}=D-\gamma_{e}B_{0}\approx2\pi\times\SI{1.47}{GHz}$. The AC magnetic field $b_{s}(t)$ then produces the signal Hamiltonian $H_{\text{sig}}(t)=-\gamma_{e}b_{s}(t)\sigma_{z}/2$, which has the same form as the signal Hamiltonian in Eq.~(\ref{eq1}).

In practice, the field $b(t)$ sensed by the NV center contains both the target signal $b_{s}(t)$ and magnetic noise $b_{n}(t)$ arising, for example, from the diamond surface, the nuclear-spin bath, or external electronics. Signals from non-target nuclear spins can be particularly problematic when their Larmor frequencies coincide with harmonics of the target frequency. For example, the magnitude of the proton gyromagnetic ratio ($|\gamma_{n}|=2\pi\times\SI{42.57}{MHz/T}$) is approximately five times that of $^{29}\mathrm{Si}$ ($|\gamma_{n}|=2\pi\times\SI{8.47}{MHz/T}$). Thus, when detecting $^{29}\mathrm{Si}$, a proton NMR signal can overlap with the fifth-harmonic response of a conventional pulse-based protocol and lead to ambiguous signal identification. Such protocols suffer from spectral leakage and non-ideal responses at harmonics of the target frequency \cite{Zeng2024,PRX2015}. In contrast, the proposed adiabatic quantum lock-in protocol employs sinusoidal modulation functions to avoid the spectral leakage inherent to pulse-based control, thereby suppressing spurious harmonic resonances and resolving this ambiguity.

Sections~\ref{AD1} and \ref{AD2} quantify this advantage through, respectively, an analytical filter-function comparison and numerical simulations in the presence of harmonic noise. The proposed protocol responds selectively at the target frequency, strongly suppresses unwanted-frequency responses, and yields a substantial SNR improvement. It therefore provides a promising approach to microscale NMR with NV-center sensors.

We emphasize that this discussion concerns regimes in which a semiclassical description of the NMR signal is applicable \cite{Laraoui11}. For a single or a few strongly coupled nuclear spins \cite{DJF2024}, probe back-action and probe-target quantum correlations become non-negligible, requiring a fully quantum treatment of the joint dynamics beyond the scope of the present work.

\subsection{Implementation scheme}\label{ISS}

To illustrate the essence of adiabatic quantum lock-in detection, in Section \ref{GP}, we focus on the case where each lock-in signal is measured using two or three successive adiabatic evolution processes occurring within a single period of the target signal. In practice, the measurement process can be repeated periodically to accumulate more relative phases, thereby enabling the detection of weak signals. Moreover, the phase $\phi$ in the Hamiltonian (\ref{eq1}) is a free parameter, which indicates that the proposed general protocol is compatible with composite pulse sequences \cite{Levitt1986,Tycko1984,Torosov2011,Genov2020,Paspalakis19}, such as XY4 and XY8. The combination of composite pulse sequences and the adiabatic quantum lock-in detection protocol can further enhance robustness against control errors. With these concepts, we propose a practical implementation scheme for adiabatic quantum lock-in detection using NV centers in diamond.

In practice, the electron spin of the NV center can be initialized into $\ket{m_{s}=0}\equiv\ket{1}$ by using optical pumping cycles \cite{Robledo11}. By applying a resonant driving microwave field with frequency $\omega_0$ that couples the two Zeeman levels $\ket{m_{s}=-1}$ and $\ket{m_{s}=0}$, the gate operations used for preparing the initial state and for readout can be implemented. The detection sequence described by the control Hamiltonian in Eq.~(\ref{eq1}) can be realized by applying a driving microwave field with time-dependent Rabi frequency $\Omega_{R}(t)$, frequency $\omega_{MW}(t)=\omega_0-\Delta(t)$, and phase $\phi$. Finally, the populations of the spin states can be measured by detecting the fluorescence emitted during the optical pumping cycles, enabling extraction of information from the final state of the evolution process.

As shown in Fig.~\ref{Fig1}(a), the entire lock-in detection process is divided into two stages: the first stage measures in-phase component $I_{\text{sig}}$, and the second stage measures quadrature component $Q_{\text{sig}}$. If the interval between the driving microwave fields used to measure $I_{\text{sig}}$ and $Q_{\text{sig}}$ equals an integer multiple of the period of the target signal, the measurement result is identical to that obtained when $I_{\text{sig}}$ and $Q_{\text{sig}}$ are measured simultaneously. Therefore, the proposed adiabatic quantum lock-in detection can be implemented using NV centers. Furthermore, simultaneous measurement of $I_{\text{sig}}$ and $Q_{\text{sig}}$ can be realized if one can address and control two NV centers concurrently.

In the first stage, the spin state of the NV center is initialized in $\ket{1}$ using a laser pulse and then prepared in $\frac{1}{\sqrt{2}}(\ket{0}+\ket{1})$ by implementing an $R_{y}(-\pi/2)$ gate. After that, the spin state of the NV center evolves under the influence of both the driving microwave field and the actual signal $M(t)$. As shown in Fig.~\ref{Fig1}(b), the XY8 sequence of the driving microwave field consists of eight segments, each of duration $T/2$. In the odd-numbered segments, the Rabi frequency and detuning of the driving field are set to $\Omega_{R}(t)=A\sin\omega t$, $\Delta(t)=A\cos\omega t$. In the even-numbered segments, the Rabi frequency and detuning of the driving field are set to $\Omega_{R}(t)=A\sin(\omega t-\pi)$, $\Delta(t)=A\cos(\omega t-\pi)$. In the segments labeled X, the phase $\phi$ of the driving field is set to $\phi=0$. In the segments labeled Y, the phase $\phi$ of the driving field is set to $\phi=\pi/2$. A single XY8 sequence lasts $4T$ and spans four periods of the target signal. According to the results presented in Section \ref{GP}, the superposition state acquires a relative phase of $8N\alpha(T,0)$ after $N$ repetitions of the XY8 sequence and evolves into $\frac{1}{\sqrt{2}}[\ket{0}+e^{i8N\alpha(T,0)}\ket{1}]$. To measure the relative phase, an $R_{x}(\pi/2)$ gate is implemented, after which the populations of states $\ket{0}$ and $\ket{1}$ are $P_{0}=\{1+\sin[8N\alpha(T,0)]\}/2$ and $P_{1}=\{1-\sin[8N\alpha(T,0)]\}/2$. Therefore, $\alpha(T,0)$ can be obtained by measuring the populations of the spin states. Furthermore, the in-phase component $I_{\text{sig}}$ is given by $I_{\text{sig}}=\alpha(T,0)/T$.

In the second stage, the spin state of the NV center is initialized into $\ket{1}$ via a laser pulse and subsequently evolves under the combined influence of the driving microwave field and the actual signal $M(t)$. As shown in Fig.~\ref{Fig1}(c), the XY8 sequence of the driving microwave field consists of seven segments of duration $T/2$ and two end segments of duration $T/4$. In the odd-numbered segments, the Rabi frequency and detuning of the driving field are set to $\Omega_{R}(t)=A\sin(\omega t+\pi/2)$, $\Delta(t)=A\cos(\omega t+\pi/2)$. In the even-numbered segments, the Rabi frequency and detuning of the driving field are set to $\Omega_{R}(t)=A\sin(\omega t-\pi/2)$, $\Delta(t)=A\cos(\omega t-\pi/2)$. In the segments labeled X, the phase $\phi$ of the driving field is set to $\phi=0$. In the segments labeled Y, the phase $\phi$ of the driving field is set to $\phi=\pi/2$. A single XY8 sequence lasts $4T$ and spans four periods of the target signal. According to the results presented in Section \ref{GP}, the propagator of the single adiabatic XY8 sequence in the basis $\{\ket{0},\ket{1}\}$ takes the form of
\begin{equation}
\begin{split}
U(4T,0)=\left(\begin{array}{ccc}
         \cos[4\beta(T,0)] & i\sin[4\beta(T,0)]\\
         i\sin[4\beta(T,0)] & \cos[4\beta(T,0)]\\
       \end{array}\right).
\end{split}
     \label{eq240}
\end{equation}
After $N$ repetitions of the XY8 sequence, the propagator becomes
\begin{equation}
\begin{split}
U(4NT,0)=\left(\begin{array}{ccc}
         \cos[4N\beta(T,0)] & i\sin[4N\beta(T,0)]\\
         i\sin[4N\beta(T,0)] & \cos[4N\beta(T,0)]\\
       \end{array}\right).
\end{split}
     \label{eq250}
\end{equation}
Therefore, after $N$ repetitions of the XY8 adiabatic evolution processes, the spin state of the NV center evolves from $\ket{1}$ to $i\sin[4N\beta(T,0)]\ket{0}+\cos[4N\beta(T,0)]\ket{1}$. To measure the parameter $\beta(T,0)$, the $R_{y}(\pi/2)$ and $R_{x}(\pi/2)$ gates are implemented in sequence. After that, the populations of the states $\ket{0}$ and $\ket{1}$ are given by $P_{0}=\{1-\sin[8N\beta(T,0)]\}/2$ and $P_{1}=\{1+\sin[8N\beta(T,0)]\}/2$. Therefore, $\beta(T,0)$ can be obtained by measuring the populations of the spin states. Furthermore, the quadrature component $Q_{\text{sig}}$ is given by $Q_{\text{sig}}=\beta(T,0)/T$.

\section{ANALYSIS AND DISCUSSION}\label{AD}

In this section, we discuss the performance of the proposed adiabatic quantum lock-in detection protocol. First, we analyze the filter function to discuss the advantages of the adiabatic quantum lock-in detection protocol. Second, we investigate the performance of the proposed protocol under the influence of different types of background noise. Finally, we discuss the robustness of the adiabatic detection sequence against control errors and decoherence. Moreover, we present some results obtained using the measurement protocol based on dynamical decoupling (DD) pulse sequences for comparison.

\begin{figure}[htb]
  \includegraphics[scale=0.06]{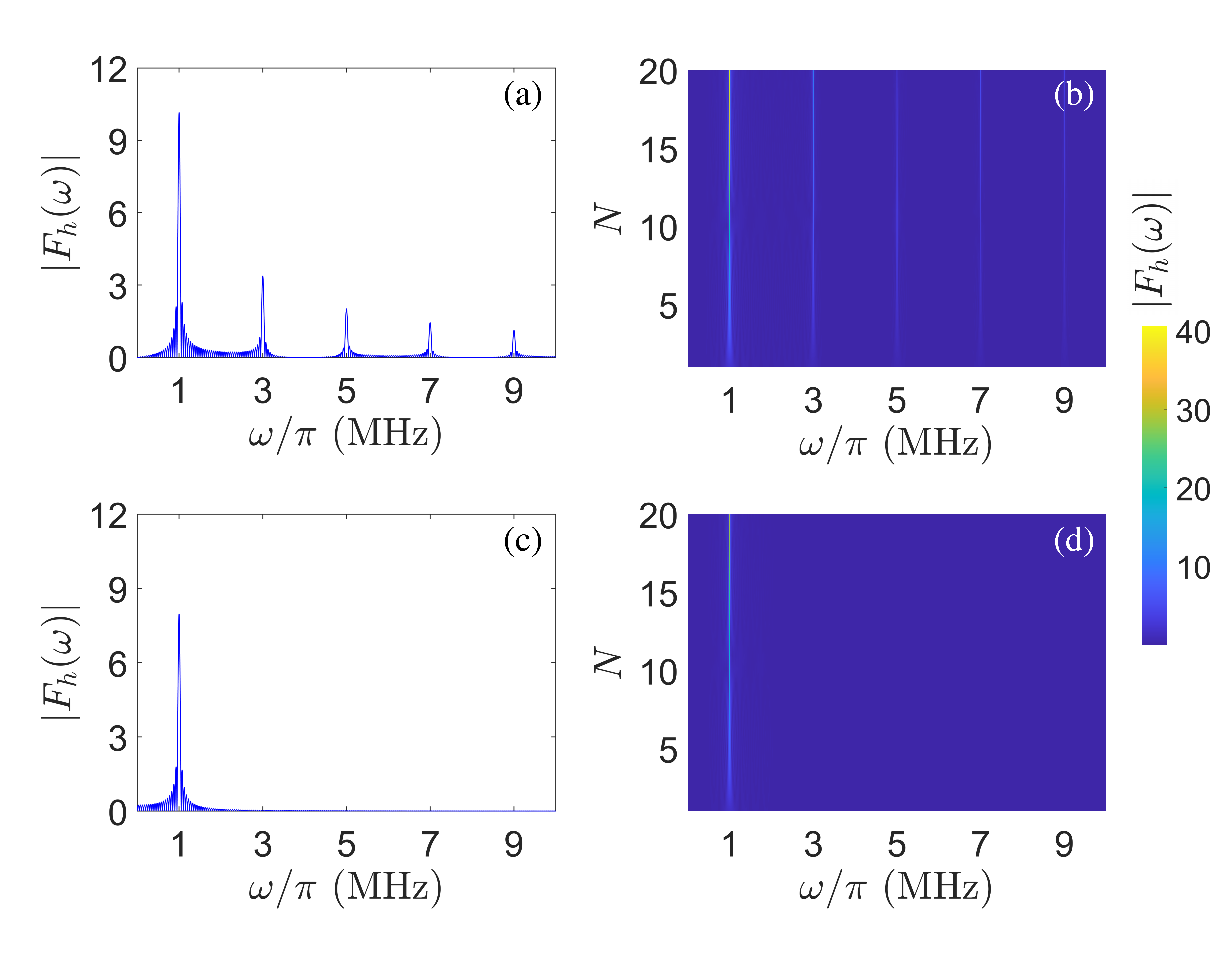}
  \caption{Comparison between the filter functions generated by the proposed adiabatic detection sequence and those generated by the conventional dynamical decoupling sequence. (a) The filter function of the conventional dynamical decoupling sequence with $N=5$ is plotted as a function of $\omega$. (b) The filter function of the conventional dynamical decoupling sequence is plotted as a function of $\omega$ and $N$. (c) The filter function generated by the proposed adiabatic detection sequence with $N=5$ is plotted as a function of $\omega$. (d) The filter function generated by the proposed adiabatic detection sequence is plotted as a function of $\omega$ and $N$. }
   \label{Fig2}
\end{figure}

\subsection{Filter function analysis}\label{AD1}

The fundamental advantage of lock-in detection lies in its capability to extract the information of the target signal from background noise. Therefore, it is necessary to analyze the quantum lock-in detection protocol in the frequency domain and quantitatively evaluate its noise filtering capability. In the following, we derive the filter function of the proposed protocol and compare it with that of the pulse-based measurement method.

As shown in Eqs.~(\ref{eq14}) and (\ref{eq22}), the response of the detection system to the external signal $M(t)$ is dependent on the phases accumulated during the detection process, which can generally be expressed as
\begin{align}
R=\int^{+\infty}_{-\infty}M(t)h(t)dt,
     \label{eq33}
\end{align}
where the time window of the detection process is incorporated into the modulation function $h(t)$. By using the Fourier transform
\begin{align}
\begin{split}
F_{f}(\omega)=\frac{1}{\sqrt{2\pi}}\int^{+\infty}_{-\infty}f(t)e^{-i\omega t}dt,\\
f(t)=\frac{1}{\sqrt{2\pi}}\int^{+\infty}_{-\infty}F_{f}(\omega)e^{i\omega t}d\omega,
\end{split}
     \label{eq34}
\end{align}
Eq.~(\ref{eq33}) can be rewritten as
\begin{align}
R=\int^{+\infty}_{-\infty}F_{M}(\omega)F_{h}^{\ast}(\omega)d\omega,
     \label{eq35}
\end{align}
where $F_{M}(\omega)$ and $F_{h}(\omega)$ are the Fourier transforms of $M(t)$ and $h(t)$, respectively.
This indicates that the response of the detection system depends on the overlap between $F_{M}(\omega)$ and $F_{h}^{\ast}(\omega)$ in the frequency domain. Therefore, the spectral bandwidth of the modulation function $h(t)$ in the frequency domain serves as a crucial indicator for assessing the frequency selectivity performance of the lock-in detection protocol. Here, we define the filter function as the modulus of  $F_{h}(\omega)$, which takes the form of
\begin{align}
|F_{h}(\omega)|=\frac{1}{\sqrt{2\pi}}\left|\int^{+\infty}_{-\infty}h(t)e^{-i\omega t}dt\right|.
     \label{eq36}
\end{align}

To illustrate the advantages of the adiabatic quantum lock-in detection protocol, we numerically calculate the filter function generated by the proposed adiabatic detection sequence and compare it with that generated by the conventional dynamical decoupling sequence. The results are shown in Fig.~\ref{Fig2}. For the adiabatic sequence used in the quantum lock-in detection protocol, the sinusoidal-wave modulation function takes the form $h(t)=\cos(\omega t)[u(t)-u(t-4NT)]$, where $u(t)$ is the unit step function. For the conventional dynamical decoupling sequence, the bipolar square wave modulation function takes the form $h(t) = \operatorname{sgn}\big[\cos(\omega t)\big]\big[u(t) - u(t - 4NT)\big]$, where $\operatorname{sgn}[\cdot]$ denotes the sign function. In the simulation, the modulation frequency is set to $\omega=2\pi\times 0.5~\text{MHz}$. It is worth noting that replacing $\cos(\omega t)$ with $\sin(\omega t)$ in the modulation functions yields a similar result, owing to the property of the Fourier transform.

As shown in Fig.~\ref{Fig2}, the filter function generated by the proposed adiabatic detection sequence exhibits a single spectral peak at the modulation frequency $\omega$, whereas the filter function generated by the conventional dynamical decoupling sequence exhibits spectral peaks at the frequencies $k\omega$, where $k\in\mathbbm{Z}^{+}_{\text{odd}}$. This indicates that the proposed adiabatic quantum lock-in detection protocol can significantly alleviate the spectral leakage effect and exhibits excellent frequency selectivity. Therefore, the adiabatic quantum lock-in detection protocol is expected to achieve enhanced robustness against broadband background noise, which is crucial for weak signal detection.

\subsection{Robustness against background noise}\label{AD2}

In the following, we discuss the robustness of the proposed adiabatic quantum lock-in detection protocol against background noise through numerical simulations. Recall that the actual signal $M(t)$ sensed by the quantum probe is the superposition of the target signal $S(t)$ and the background noise $N(t)$, see Eq.~(\ref{eq_signalwithnoise}). In practice, the ability to filter out background noise and extract information about the target signal is one of the most important aspects of the lock-in detection protocol. Here, we consider two prevalent noise models, white noise and harmonic noise, and simulate the performance of our protocol.

\begin{figure}[tb]
  \includegraphics[scale=0.6]{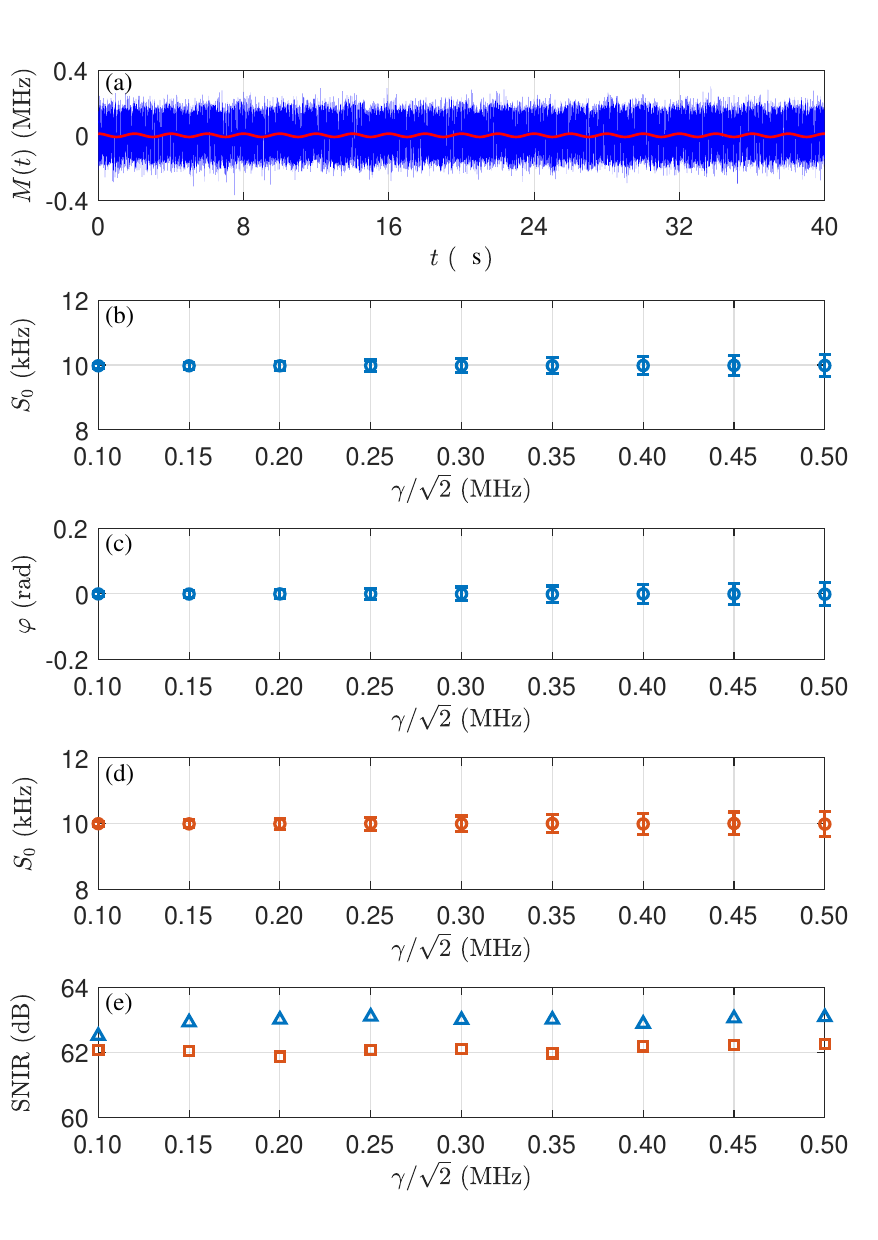}
  \caption{The performance of the detection protocols under different white noise intensities. (a) The noise-free signal (red line) and the noisy signal with $\gamma = 0.1/\sqrt{2}~\text{MHz}$ (blue line). (b) The means and standard deviations of the target signal amplitude measurement results obtained using the adiabatic quantum lock-in detection protocol. (c) The means and standard deviations of the target signal phase measurement results obtained using the adiabatic quantum lock-in detection protocol. (d) The means and standard deviations of the target signal amplitude measurement results obtained using the DD sequence. (e) The SNIR values of the adiabatic quantum lock-in detection protocol (blue triangles) and the DD-based protocol (orange squares).}
   \label{Fig3}
\end{figure}

In the field of weak signal detection, white noise is the most prevalent and dominant type of background noise \cite{wnoise}. White noise is a random signal characterized by a constant power spectral density across the entire frequency domain. To investigate the performance of the lock-in detection protocol under the influence of white noise, we numerically simulate the evolution of the quantum probe governed by the Hamiltonian
\begin{align}
\begin{split}
H(t)=H_{\text{ctrl}}(t)+[S_{0}\cos(\omega_{c}t+\varphi)+N(t)]\sigma_{z},
\end{split}
\label{eqHE}
\end{align}
where $N(t)$ represents the white noise. In the simulation, the target signal has an amplitude of $S_{0} = 10\ \text{kHz}$, an initial phase of $\varphi=0$, and a carrier frequency of $\omega_c = 2\pi \times 0.5\ \text{MHz}$. The control Hamiltonian $H_{\text{ctrl}}(t)$ is set according to the implementation scheme in Section \ref{IS}. The peak values of the Rabi frequency and the detuning of the control field are both set to $A = 2\pi \times 10~\text{MHz}$. The number of repetitions of the adiabatic XY8 sequence is set to $N=5$, and the total duration of the detection process is \SI{40}{\micro\second}. The white noise $N(t)$ is generated from a normal distribution with zero mean and standard deviation $\gamma$. In this case, the power of the target signal is $P_{S}=S_{0}^{2}/2$, and the power of the white noise is $P_{N}=\gamma^{2}$. The signal-to-noise ratio (SNR) of the actual input signal sensed by the quantum probe can be calculated by
\begin{align}
\begin{split}
\text{SNR}_{\mathrm{in}}=10\log_{10}(P_{S}/P_{N}),
\end{split}
\label{SNR1}
\end{align}
in units of decibel (dB). Similarly, the SNR of the measurement result can be defined as
\begin{align}
\begin{split}
\text{SNR}_{\mathrm{out}}=10\log_{10}(S_{0}^{2}/\langle(\tilde{S}_{0}-S_{0})^{2}\rangle),
\end{split}
\label{SNR2}
\end{align}
where $S_{0}$ denotes the true amplitude of the target signal, $\tilde{S}_{0}$ denotes the measured amplitude, and $\langle\cdot\rangle$ denotes the statistical average over multiple measurements. The signal-to-noise improvement ratio (SNIR) can be further defined as
\begin{align}
\begin{split}
\mathrm{SNIR}=10\log_{10}\left(\frac{S_{0}^{2}/\langle(\tilde{S}_{0}-S_{0})^{2}\rangle}{P_{S}/P_{N}}\right).
\end{split}
\label{SNIR}
\end{align}

We simulate the performance of the proposed adiabatic quantum lock-in detection protocol under different noise intensities. The mean
and standard deviation of the measurement results for amplitude and phase were obtained from 2500 numerical simulations of the detection process. Moreover, we also simulate the performance of the conventional measurement method based on the DD sequence (XY8), which can measure the amplitude of ac fields under the in-phase condition $\varphi=0$. For the XY8 sequence, the Rabi frequency is set to $\Omega_{\text{max}}=2\pi \times 10~\text{MHz}$, the duration of each pulse is \SI{0.05}{\micro\second}, the interval between pulses is \SI{0.95}{\micro\second}, and the duration of single XY8 sequence is \SI{8}{\micro\second}. The XY8 sequence is also repeated five times. The simulation results are shown in Fig.~\ref{Fig3}.

In Fig.~\ref{Fig3}(a), the waveforms of the noise-free signal and a realization of the noisy signal with $\gamma = 0.1/\sqrt{2}~\text{MHz}$ are presented. In Fig.~\ref{Fig3}(b)–(d), the means and standard deviations of the measurement results are plotted as functions of the white noise intensity. For the adiabatic quantum lock-in detection protocol, the amplitude and phase of the target signal are reconstructed from the in-phase and quadrature measurements. For the DD-based protocol, the amplitude of the target signal is measured under the in-phase condition. For both protocols, the measurement results follow a probability distribution centered on the true value. The average value of the measurement results converges to the true value, and the standard deviation of the measurement results increases as the white noise intensity increases. In Fig.~\ref{Fig3}(e), the SNIR values of the detection protocols are plotted as functions of the noise intensity. As anticipated, the SNIR values of our adiabatic quantum lock-in detection protocol are always higher than those of the DD-based protocol. This enhancement results from the sinusoidal-wave modulation, which alleviates spectral leakage and exhibits excellent frequency selectivity. The advantage of the adiabatic quantum lock-in detection protocol becomes even more evident when considering noise with a specific spectral distribution.

We now  proceed to investigate the performance of the detection protocols under the influence of harmonic noise, which is a type of noise commonly present in electrical equipment \cite{hnoise} and can cause ambiguities in NMR spectroscopy \cite{PRX2015}. In this case, the noise term in Eq.~(\ref{eqHE}) can be expressed as
\begin{align}
\begin{split}
N(t)=\sum^{+\infty}_{n=2}N_{n}\cos(n\omega_{c}t+\varphi_{n}),
\end{split}
     \label{hnoise}
\end{align}
where $N_{n}$ and $\varphi_{n}$ are the amplitudes and phases of the $n$-th harmonic. In this case, the power of the harmonic noise is $P_{N}=(\sum_{n}N_{n}^{2})/2$. Here, we consider only the second and third harmonic noise and set $N_2 = N_3 = \gamma$. The phases $\varphi_{2}$ and $\varphi_{3}$ are randomly generated in each realization of the simulation. The SNIR can also be calculated using Eq.~(\ref{SNIR}).

\begin{figure}[tb]
  \includegraphics[scale=0.6]{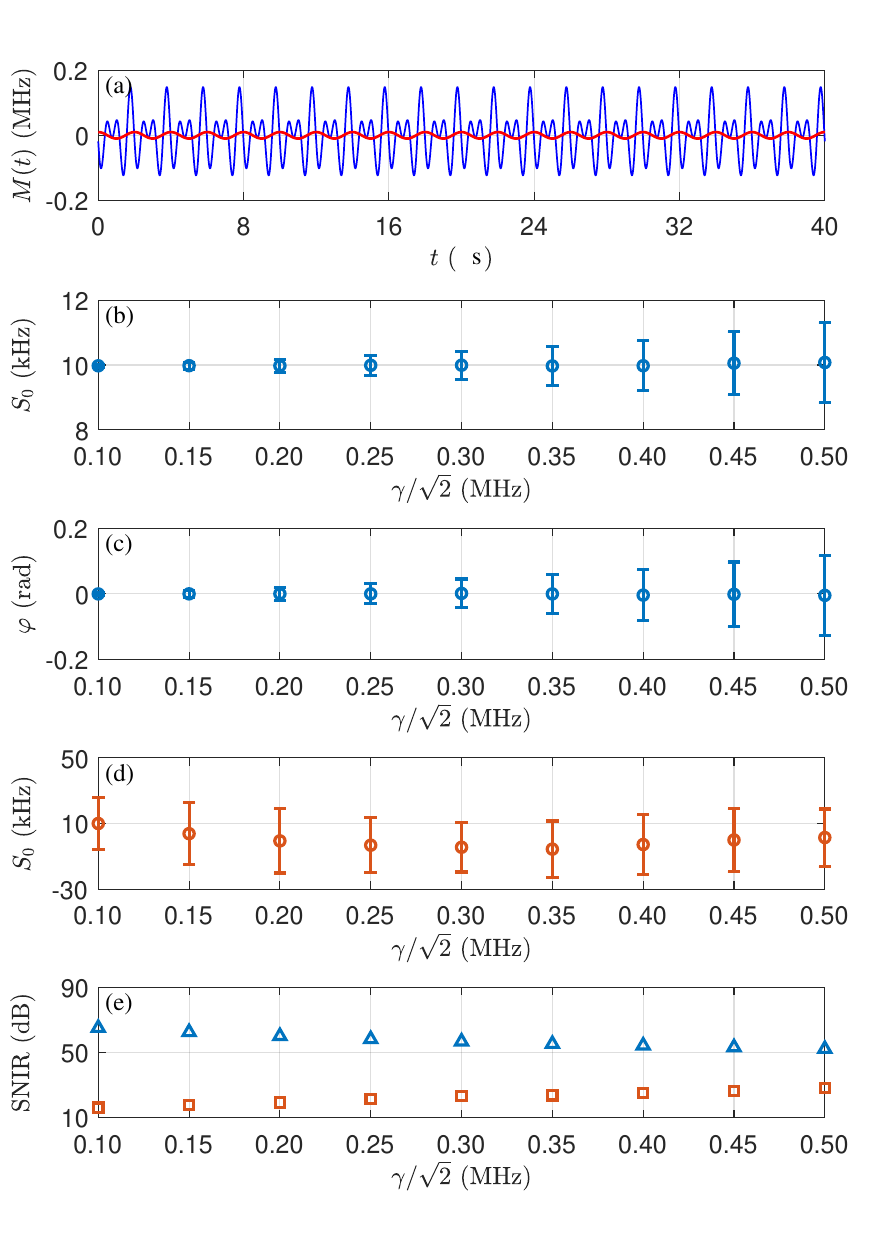}
  \caption{The performance of the detection protocols under different harmonic noise intensities. (a) The noise-free signal (red line) and the noisy signal with $\gamma = 0.1/\sqrt{2}~\text{MHz}$ (blue line). (b) The means and standard deviations of the target signal amplitude measurement results obtained using the adiabatic quantum lock-in detection protocol. (c) The means and standard deviations of the target signal phase measurement results obtained using the adiabatic quantum lock-in detection protocol. (d) The means and standard deviations of the target signal amplitude measurement results obtained using the DD sequence. (e) The SNIR values of the adiabatic quantum lock-in detection protocol (blue triangles) and the DD-based protocol (orange squares).}
   \label{Fig4}
\end{figure}

The simulation results are shown in Fig.~\ref{Fig4}. In Fig.~\ref{Fig4}(a), the waveforms of the noise-free signal and a realization of the noisy signal with $\gamma = 0.1/\sqrt{2}~\text{MHz}$ are presented. In Fig.~\ref{Fig4}(b)–(d), the means and standard deviations of the measurement results are plotted as functions of the harmonic noise intensity. For the adiabatic quantum lock-in detection protocol, the average values of the measured amplitude and phase of the target signal converge to their true values, while the standard deviation of the measurements increases with increasing noise intensity. For the DD-based protocol, the average value of the amplitude measurement results deviates from the true value, indicating that the measurement is severely impaired by harmonic noise. In Fig.~\ref{Fig4}(e), the SNIR values of the detection protocols are plotted as functions of the noise intensity, showing that the adiabatic quantum lock-in detection protocol can significantly improve the SNR even under extreme harmonic noise. These results indicate that the adiabatic quantum lock-in detection protocol can significantly suppress the influence of harmonic noise, which could instead lead to the failure of conventional DD-based protocols.

The analysis presented in this section confirms that our proposed adiabatic quantum lock-in detection protocol can effectively and accurately  extract information about the target signal, maintaining a remarkably high SNR, even when the target signal is submerged in extreme background noise.

\subsection{Robustness against control errors}

In practical realizations, inevitable control errors arising from imperfect experimental equipment may seriously disrupt detection processes \cite{Poggi24}. Therefore, we demonstrate the robustness of the detection sequences against control errors by simulating their performance in the absence of input signals. For the adiabatic control sequences (XY8 and repeated X) proposed for the quantum lock-in detection protocol, the peak values of the Rabi frequency and the detuning are both set to $\Omega_{\text{max}}=A=2\pi \times 10~\text{MHz}$, the frequency of the target signal is set to $\omega_{c}=2\pi\times 0.5~\text{MHz}$, and the total duration of each sequence is \SI{8}{\micro\second}. For the DD sequences (XY8 and CPMG), the Rabi frequency is set to $\Omega_{\text{max}}=2\pi \times 10~\text{MHz}$, the duration of each pulse is \SI{0.05}{\micro\second}, the interval between pulses is \SI{0.95}{\micro\second}, and the total duration of each sequence is \SI{8}{\micro\second}.

To investigate the robustness of the control sequences, we numerically simulate their performance under the influence of control errors. The Rabi frequency error is introduced by varying the Rabi frequency from $\Omega$ to $(1+\epsilon)\Omega$, and the detuning error is introduced by adding $(\delta\Omega_{\text{max}}/2)\sigma_{z}$ to the Hamiltonian in Eq.~(\ref{eq1}). We utilize the infidelity $1-F$ to assess the performance of the sequences, where the fidelity $F$ is defined as
\begin{equation}
F=\frac{1}{2}\left|\text{Tr}\left[U(\delta,\epsilon)U^{\dag}_{\text{ideal}}\right]\right|,
     \label{eq}
\end{equation}
in which $U_{\text{ideal}}=I$ represents the ideal evolution in the absence of control errors.

\begin{figure}[tb]
  \includegraphics[scale=0.06]{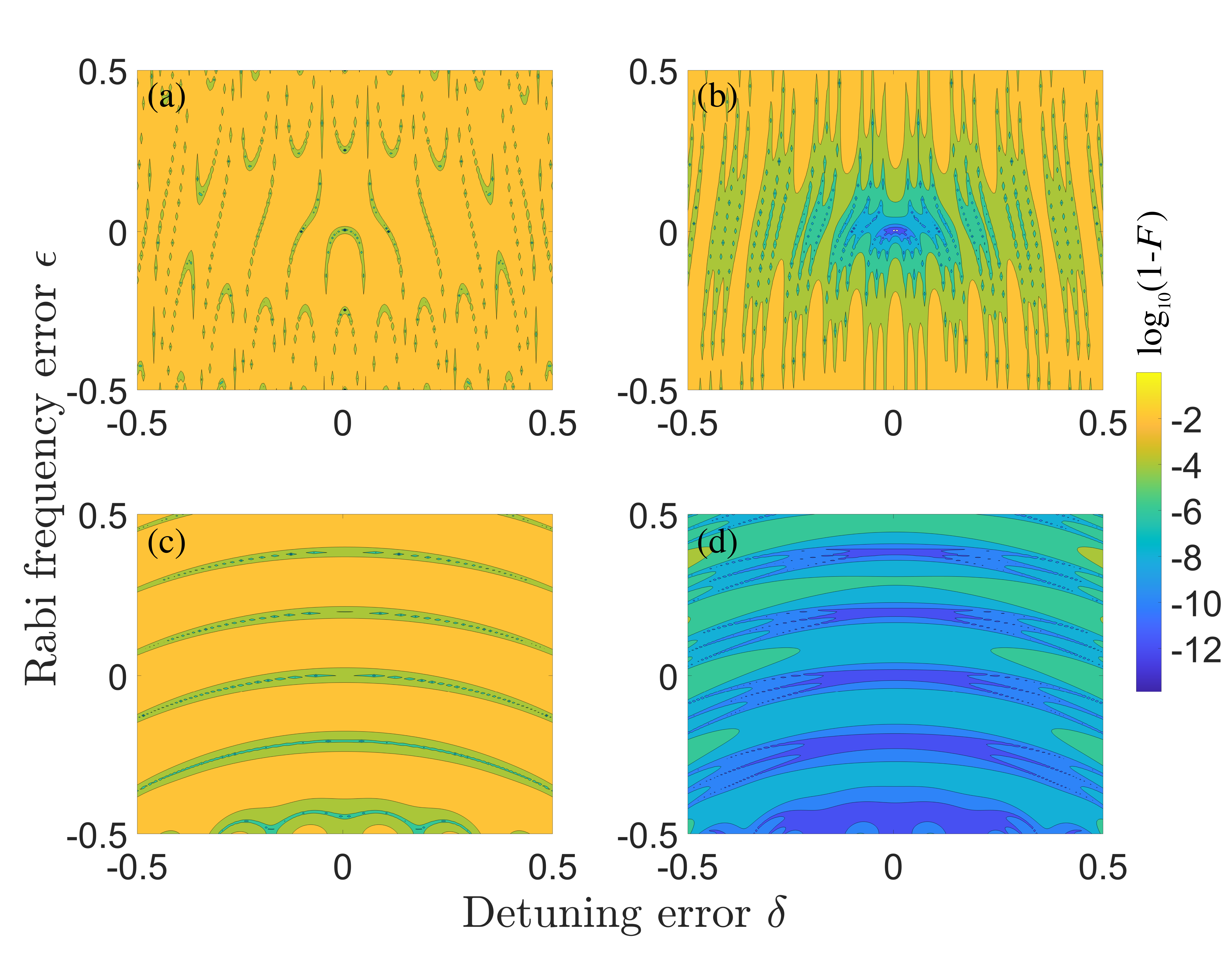}
  \caption{Performance of the control sequences under the influence of control errors. The infidelities $1-F$ of the control sequences are plotted as functions of the Rabi frequency error strength $\epsilon$ and the detuning error strength $\delta$ on a logarithmic scale. (a) The CPMG sequence with eight rectangular pulses. (b) The XY8 sequence with rectangular pulses. (c) The sequence consisting of eight repeated adiabatic evolution processes. (d) The adiabatic XY8 sequence used for quantum lock-in detection.}
   \label{Fig5}
\end{figure}

As shown in Fig.~\ref{Fig5}, the infidelities $1-F$ of the control sequences are plotted as functions of the Rabi frequency error strength $\epsilon$ and the detuning error strength $\delta$, on a logarithmic scale. The CPMG sequence, which consists of simple rectangular pulses, is sensitive to control errors and can only achieve low infidelity when the errors are extremely weak. The robustness of the sequence can be further enhanced through the combination of composite pulse techniques. The XY8 sequence composed of rectangular pulses exhibits better robustness against control errors than the CPMG sequence. Similarly, the adiabatic XY8 sequence exhibits better robustness against control errors than the sequence consisting of repeated adiabatic evolutions. Overall, the adiabatic XY8 sequence possesses the strongest robustness and can achieve an extremely low infidelity level even in the presence of significant control errors. Moreover, regardless of whether composite pulse techniques are applied, adiabatic evolution is always more robust than evolution driven by rectangular pulses. These results indicate that the proposed quantum lock-in detection protocol is capable of resisting the influence of control errors and effectively extracting the information of the target signal.

\subsection{Robustness against decoherence}

To thoroughly evaluate the performance of the detection sequences, we consider not only the control errors arising from imperfect experimental equipment but also the decoherence resulting from the inevitable interaction between NV centers and their environment. Specifically, dephasing resulting from random magnetic field fluctuations is an important source of coherence loss. In practice, random magnetic field fluctuations may arise from hyperfine interactions with the surrounding nuclear spin bath \cite{Rong14} or from inhomogeneous broadening in ensembles of NV centers \cite{Genov20}.

\begin{figure}[tb]
  \includegraphics[scale=0.62]{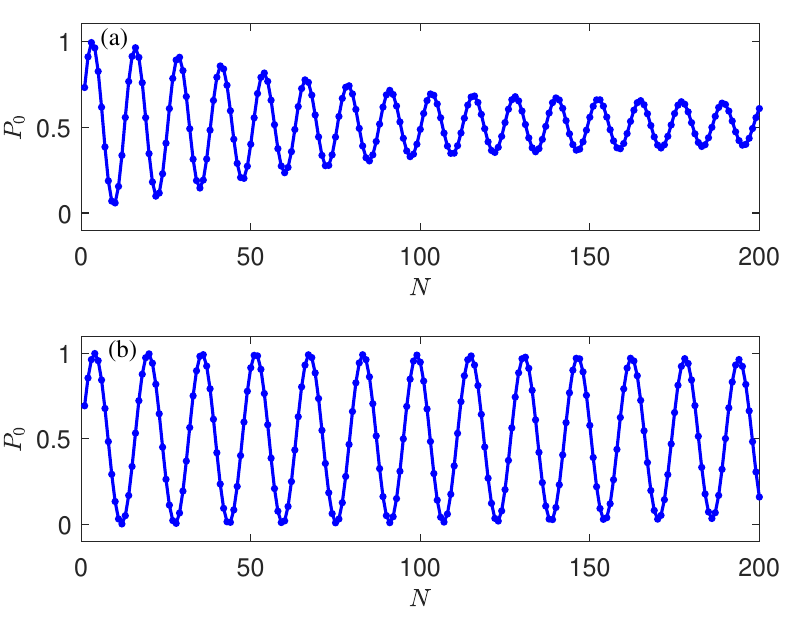}
  \caption{Performance of the detection sequences under the influence of random magnetic field fluctuations. The population $P_{0}$ of the state $\ket{0}$ is plotted as a function of the number $N$ of repetitions of each sequence. (a) The XY8 sequence with rectangular pulses. (b) The adiabatic XY8 sequence used for quantum lock-in detection.}
   \label{Fig6}
\end{figure}

Due to the presence of random magnetic field fluctuations, the system Hamiltonian includes a noise term $H^{\prime}=\delta_{0}\sigma_{z}$, in which $\delta_{0}$ depends on the strength of the random magnetic field. Here, we assume that $\delta_{0}$ follows a Gaussian distribution $P(\delta_{0})=\exp(-\delta_{0}^{2}/2\sigma^{2})/(\sigma\sqrt{2\pi})$, where $\sigma$ is the standard deviation of the distribution and we here take a standard deviation $\sigma=2\pi$ MHz. In order to investigate the influence of the dephasing effect on the detection process, we simulate the evolution of the quantum probe under both the rectangular-pulse-based XY8 sequence and the adiabatic XY8 sequence in the presence of the target signal. In the simulation, the target signal has an amplitude of $S_{0} = 50\ \text{kHz}$, an initial phase of $\varphi = 0$, and a carrier frequency of $\omega_c = 2\pi \times 0.5\ \text{MHz}$. The parameters of the detection sequences are the same as those in the previous section. The initial state of the quantum probe is set to $\frac{1}{\sqrt{2}}(\ket{0}+\ket{1})$, and then $N$ repetitions of the detection sequence are applied. Following the evolution, an $R_{x}(\pi/2)$ gate is applied, and ultimately, the population of the state $\ket{0}$ is calculated. By performing the simulation 2500 times for different noise realizations and averaging the outcomes, we can obtain the results shown in Fig.~\ref{Fig6}.

As shown in Fig.~\ref{Fig6}, for the rectangular-pulse-based XY8 sequence, the coherence of the quantum probe decays as the number of repetitions of the sequence increases, which is caused by the distribution of the random magnetic field. In contrast, the coherence of the quantum probe driven by the adiabatic XY8 sequence decays only negligibly. This indicates that the adiabatic XY8 sequence can protect the coherence of the quantum probe more effectively, thereby extending the coherence time. A longer coherence time allows for a longer sensing duration and a greater number of repetitions of the detection sequence, which is of great significance for enhancing sensitivity.

\subsection{Precision analysis}

Finally, we analyze the precision limits of the proposed adiabatic lock-in protocol. Specifically, we consider the implementation scheme based on the XY8 sequence presented in Sec.~\ref{ISS}.

Under the condition $\omega=\omega_{c}$ and in the absence of noise $N(t)$, the target signal in Eq.~(\ref{eq_signal}) can be reexpressed in terms of its in-phase and quadrature components as

\begin{align}
S(t)=2I_{\text{sig}}\cos(\omega_{c} t)+2Q_{\text{sig}}\sin(\omega_{c} t).
\label{eq_signal2}
\end{align}
In the implementation scheme, the two components $I_{\text{sig}}$ and $Q_{\text{sig}}$ are measured separately. First, we discuss the precision limits for estimating $I_{\text{sig}}$ and $Q_{\text{sig}}$ separately.

For measuring the in-phase component $I_{\text{sig}}$, the XY8 sequence shown in Fig.~\ref{Fig1}(b) has a duration of $4T$ and spans four periods of the target signal. We assume that the quantum probe is prepared in $\frac{1}{\sqrt{2}}(\ket{0}+\ket{1})$ and that the XY8 sequence is repeated $N$ times. According to the results presented in Section~\ref{GPB}, the superposition state acquires a relative phase of $8N\alpha(T,0)$ after $N$ repetitions of the XY8 sequence and evolves into

\begin{align}
\ket{\psi(4NT)}=\frac{1}{\sqrt{2}}\left[\ket{0}+e^{i8N\alpha(T,0)}\ket{1}\right],\label{eq46}
\end{align}
in which $\alpha(T,0)=\int^{T}_{0}M(t)\cos(\omega t)dt$. Under the condition $\omega=\omega_{c}$ and in the absence of noise $N(t)$, we obtain $\alpha(T,0)=TI_{\text{sig}}$. The quantum Fisher information (QFI) for estimating $I_{\text{sig}}$ can then be calculated from the quantum state in Eq.~(\ref{eq46}) as
\begin{align}
F(I_{\text{sig}})=4\left[
\langle\partial_{I_{\text{sig}}}\psi|\partial_{I_{\text{sig}}}\psi\rangle-\left|\langle\psi|\partial_{I_{\text{sig}}}\psi\rangle\right|^2
\right]=64N^{2}T^{2}.
\end{align}
The precision of estimating $I_{\text{sig}}$ is then bounded by the quantum Cramér-Rao bound \cite{Paris},
\begin{align}
\Delta I_{\text{sig}}\geq\frac{1}{\sqrt{F(I_{\text{sig}})}}=\frac{1}{8NT},\label{eq48}
\end{align}

For measuring the quadrature component $Q_{\text{sig}}$, the XY8 sequence shown in Fig.~\ref{Fig1}(c) has a duration of $4T$ and spans four periods of the target signal. We assume that the quantum probe is prepared in $\ket{1}$ and that the XY8 sequence is repeated $N$ times. Therefore, after $N$ repetitions of the XY8 adiabatic evolution process, the state of the probe evolves from $\ket{1}$ to
\begin{align}
\ket{\psi(4NT)}=i\sin[4N\beta(T,0)]\ket{0}+\cos[4N\beta(T,0)]\ket{1},\label{eq49}
\end{align}
in which $\beta(T,0)=\int^{T}_{0}M(t)\sin(\omega t)dt$. Under the condition $\omega=\omega_{c}$ and in the absence of noise $N(t)$, we obtain $\beta(T,0)=TQ_{\text{sig}}$.
The QFI for estimating $Q_{\text{sig}}$ can then be calculated from the quantum state in Eq.~(\ref{eq49}) as

\begin{align}
F(Q_{\text{sig}})=4\left[
\langle\partial_{Q_{\text{sig}}}\psi|\partial_{Q_{\text{sig}}}\psi\rangle-\left|\langle\psi|\partial_{Q_{\text{sig}}}\psi\rangle\right|^2
\right]=64N^{2}T^{2}.
\end{align}
The precision limit for estimating $Q_{\text{sig}}$ is given by
\begin{align}
\Delta Q_{\text{sig}}\geq\frac{1}{\sqrt{F(Q_{\text{sig}})}}=\frac{1}{8NT}.\label{eq51}
\end{align}

Based on the precision limits for estimating $I_{\text{sig}}$ and $Q_{\text{sig}}$, we can analyze the precision limits for estimating the amplitude $S_{0}$ and phase $\varphi$ of the target signal using the standard error propagation formula. From the relation $S_{0}=2(I_{\text{sig}}^{2}+Q_{\text{sig}}^{2})^{1/2}$, we can obtain
\begin{equation}
\Delta S_0=\frac{2
\sqrt{
I_{\text{sig}}^2(\Delta I_{\text{sig}})^2
+
Q_{\text{sig}}^2(\Delta Q_{\text{sig}})^2
}}
{\sqrt{I_{\text{sig}}^2+Q_{\text{sig}}^2}}.
\end{equation}
By using Eqs.~(\ref{eq48}) and (\ref{eq51}), the precision limit for estimating $S_{0}$ is given by
\begin{equation}
\Delta S_{0}
\geq
\frac{1}{4NT}.
\end{equation}
From the relation $\varphi=\arctan(-Q_{\text{sig}}/I_{\text{sig}})$, we can obtain
\begin{equation}
\Delta\varphi=
\frac{
\sqrt{
Q_{\text{sig}}^2(\Delta I_{\text{sig}})^2
+
I_{\text{sig}}^2(\Delta Q_{\text{sig}})^2
}}
{I_{\text{sig}}^2+Q_{\text{sig}}^2}.
\end{equation}
By using Eqs.~(\ref{eq48}) and (\ref{eq51}), the precision limit for estimating $\varphi$ is given by

\begin{equation}
\Delta\varphi
\geq
\frac{1}
{8NT\sqrt{I_{\text{sig}}^2+Q_{\text{sig}}^2}}
=
\frac{1}{4NT S_0}.
\end{equation}
Since each XY8 sequence lasts $4T$, these bounds show that, in the ideal noise-free setting, the estimation uncertainties decrease inversely with the total coherent interrogation time, while the phase uncertainty also decreases with increasing signal amplitude.

\section{CONCLUSION}\label{CON}

In conclusion, we propose a general protocol for realizing quantum lock-in detection by employing successive quantum adiabatic evolution. In our protocol, the signal modulation is achieved by adiabatically controlling the time evolution of the quantum probe, which enables the implementation of sinusoidal modulation functions in the quantum regime. The realization of sinusoidal-wave modulation substantially mitigates the problem of spectral leakage and facilitates the extraction of the complete characteristics of the target signals. Extensive simulations demonstrate that our protocol has excellent frequency selectivity, which enables a significant improvement in the signal-to-noise ratio (SNR) over existing protocols. Moreover, benefiting from composite adiabatic evolution, our protocol is shown to exhibit remarkable robustness against experimental imperfections and decoherence effects.

We further present a practical implementation scheme of our adiabatic quantum lock-in detection based on NV centers in diamond, which demonstrates that our protocol is feasible under current experimental conditions. Our proposal therefore has the potential to become a central component of quantum sensing applications. The proposed technology opens in fact a promising route for noise-resistant detection of weak alternating signals with high spatial resolution, which could shed light on NMR spectroscopy \cite{Aslam17,Munuera17}, microwave-field characterization \cite{Wang22,Meinel21}, and noise spectroscopy detection \cite{Bylander11,Wudarski23}.

Meanwhile, our protocol is not limited to NV centers and can be realized using any controllable two-level quantum system. This level of generality enables implementation on various physical platforms, such as Rydberg atoms \cite{RB20,ZHANGLJ2024}, trapped ions \cite{Nature2011,TI21}, and superconducting qubits \cite{SC23}. This further indicates that the proposed quantum lock-in detection method offers not only unprecedented performance, but also has broad application prospects. Furthermore, extending the adiabatic quantum lock-in detection to multi-level probes \cite{ML17,ML26} and many-body quantum systems \cite{Mihailescu25,PRX2021} is an interesting topic for future research.

\medskip
\begin{acknowledgments}
\noindent\begin{minipage}{\columnwidth}
\hspace*{1em}
K.Z.L. gratefully acknowledges support from the National Natural Science Foundation of China (Grant No.~62405210) and the Fundamental Research Program of Shanxi Province (Grant No.~202403021212252). J.Z.T. gratefully acknowledges support from the National Natural Science Foundation of China (Grant No.~62305241) and the Fundamental Research Program of Shanxi Province (Grant No.~202203021222113). L.T.X. gratefully acknowledges support from the National Natural Science Foundation of China (Grants No.~62127817 and No.~U23A20380) and the Science and Technology Major Special Project of Shanxi Province (Grant No.~202201010101005). G.A. gratefully acknowledges support from the Engineering and Physical Sciences Research Council (EPSRC Grant No.~EP/X010929/1).
\end{minipage}
\medskip
\end{acknowledgments}

\appendix

\section{DETAILED DERIVATIONS}
In this appendix, we provide detailed derivations of the measurement procedures presented in Secs. \ref{GPB} and \ref{Q}.

\subsection{Detailed derivations of Sec. II B}
The measurement procedure for the in-phase component $I_{\text{sig}}$ presented in Sec. \ref{GPB} is divided into two stages. The first stage occurs during the time interval $t\in{[0,\tau]}$, and the second stage occurs during the time interval $t\in{(\tau,T]}$, where $\tau=T/2$.

In the first stage $t\in{[0,\tau]}$, the Rabi frequency, detuning, and phase of the control field are set as $\Omega_{R}(t)=A\sin(\omega t)$, $\Delta(t)=A\cos(\omega t)$, and $\phi=\phi_{1}$, which corresponds to the mixing angle $\theta(t)=\omega t$, where $\omega=\pi/\tau$. During this stage, the mixing angle $\theta(t)$ varies from $0$ to $\pi$. By using Eq.~(\ref{eq6}), the unitary operators $V(0,0)$ and $V(\tau,0)$ can be obtained as
\begin{align}
\begin{split}
V(0,0)=\left(
       \begin{array}{ccc}
         1 &0\\
         0 &1\\
       \end{array}
     \right),
V(\tau,0)=\left(
       \begin{array}{ccc}
         0 &-e^{i(\frac{A\tau}{2}-\phi_{1})}\\
         e^{i(-\frac{A\tau}{2}+\phi_{1})} &0\\
       \end{array}
     \right).
\end{split}
     \label{eqa1}
\end{align}
The approximate propagator of the first stage in the rotating frame is given by Eq.~(\ref{eq9}). By using Eq.~(\ref{eq5}), the propagator of the first stage in the basis $\{\ket{0},\ket{1}\}$ can be obtained as
\begin{align}
\begin{split}
U(\tau,0)&=V(\tau,0)U_{\text{rot}}^{m}(\tau,0)V^{\dag}(0,0)\\
&=\left(
       \begin{array}{ccc}
         0 &-e^{i[\frac{A\tau}{2}-\phi_{1}+\alpha(\tau,0)]}\\
         e^{i[-\frac{A\tau}{2}+\phi_{1}-\alpha(\tau,0)]} & 0\\
       \end{array}
     \right),
\end{split}
     \label{eqa2}
\end{align}
in which $\alpha(\tau,0)=\int^{\tau}_{0}M(t)\cos(\omega t)dt$.

In the second stage $t\in{(\tau,T]}$, the Rabi frequency, detuning, and phase of the control field are set as $\Omega_{R}(t)=A\sin(\omega t-\pi)$, $\Delta(t)=A\cos(\omega t-\pi)$, and $\phi=\phi_{2}$, which corresponds to the mixing angle $\theta(t)=\omega t-\pi$. During this stage, the mixing angle $\theta(t)$ varies from $0$ to $\pi$. By using Eq.~(\ref{eq6}), the unitary operators $V(\tau,\tau)$ and $V(T,\tau)$ can be obtained as
\begin{align}
\begin{split}
V(\tau,\tau)=\left(
       \begin{array}{ccc}
         1 &0\\
         0 &1\\
       \end{array}
     \right),
V(T,\tau)=\left(
       \begin{array}{ccc}
         0 &-e^{i(\frac{A\tau}{2}-\phi_{2})}\\
         e^{i(-\frac{A\tau}{2}+\phi_{2})} &0\\
       \end{array}
     \right).
\end{split}
     \label{eqa3}
\end{align}
The approximate propagator of the second stage in the rotating frame is given by Eq.~(\ref{eq11}). By using Eq.~(\ref{eq5}), the propagator of the second stage in the basis $\{\ket{0},\ket{1}\}$ can be obtained as
\begin{align}
\begin{split}
U(T,\tau)&=V(T,\tau)U_{\text{rot}}^{m}(T,\tau)V^{\dag}(\tau,\tau)\\
&=\left(
       \begin{array}{ccc}
         0 &-e^{i[\frac{A\tau}{2}-\phi_{2}-\alpha(T,\tau)]}\\
         e^{i[-\frac{A\tau}{2}+\phi_{2}+\alpha(T,\tau)]} & 0\\
       \end{array}
     \right)
\end{split}
     \label{eqa4}
\end{align}
in which $\alpha(T,\tau)=\int^{T}_{\tau}M(t)\cos(\omega t)dt$.

By substituting Eqs.~(\ref{eqa2}) and (\ref{eqa4}) into the relation $U(T,0)=U(T,\tau)U(\tau,0)$, we obtain the propagator describing the entire measurement procedure for the in-phase component, as given in Eq.~(\ref{eq13}).

\subsection{Detailed derivations of Sec. II C}
The measurement procedure for the quadrature component $Q_{\text{sig}}$ presented in Sec. \ref{Q} is divided into three stages. The first stage occurs during the time interval $t\in{[0,\tau/2]}$, the second stage occurs during the time interval $t\in{(\tau/2,3\tau/2)}$, and the third stage occurs during the time interval $t\in{[3\tau/2,T]}$, where $\tau=T/2$.

In the first stage $t\in{[0,\tau/2]}$, the Rabi frequency, detuning, and phase of the control field are set as $\Omega_{R}(t)=A\sin(\omega t+\pi/2)$, $\Delta(t)=A\cos(\omega t+\pi/2)$, and $\phi=\phi_{1}$, which corresponds to the mixing angle $\theta(t)=\omega t+\pi/2$, where $\omega=\pi/\tau$. During this stage, the mixing angle $\theta(t)$ varies from $\pi/2$ to $\pi$. By using Eq.~(\ref{eq6}), the unitary operators $V(0,0)$ and $V(\tau/2,0)$ can be obtained as
\begin{align}
\begin{split}
&V(0,0)=\left(
       \begin{array}{ccc}
         \frac{\sqrt{2}}{2} &-\frac{\sqrt{2}}{2}e^{-i\phi_{1}}\\
         \frac{\sqrt{2}}{2}e^{i\phi_{1}} &\frac{\sqrt{2}}{2}\\
       \end{array}
     \right),\\
&V\left(\frac{\tau}{2},0\right)=\left(
       \begin{array}{ccc}
         0 &-e^{i(\frac{A\tau}{4}-\phi_{1})}\\
         e^{i(-\frac{A\tau}{4}+\phi_{1})} &0\\
       \end{array}
     \right).
\end{split}
     \label{eqa5}
\end{align}
The approximate propagator of the first stage in the rotating frame is given by Eq.~(\ref{eq15}). By using Eq.~(\ref{eq5}), the propagator of the first stage in the basis $\{\ket{0},\ket{1}\}$ can be obtained as
\begin{align}
\begin{split}
U\left(\frac{\tau}{2},0\right)&=V\left(\frac{\tau}{2},0\right)U_{\text{rot}}^{m}\left(\frac{\tau}{2},0\right)V^{\dag}\left(0,0\right)\\
&=\left(
       \begin{array}{ccc}
         \frac{\sqrt{2}}{2}e^{i[\frac{A\tau}{4}-\beta(\tau/2,0)]} &-\frac{\sqrt{2}}{2}e^{i[\frac{A\tau}{4}-\phi_{1}-\beta(\tau/2,0)]}\\
         \frac{\sqrt{2}}{2}e^{i[-\frac{A\tau}{4}+\phi_{1}+\beta(\tau/2,0)]} &\frac{\sqrt{2}}{2}e^{i[-\frac{A\tau}{4}+\beta(\tau/2,0)]}\\
       \end{array}
     \right),
\end{split}
     \label{eqa6}
\end{align}
in which $\beta(\tau/2,0)=\int^{\tau/2}_{0}M(t)\sin(\omega t)dt$.

In the second stage $t\in{(\tau/2,3\tau/2)}$, the Rabi frequency, detuning, and phase of the control field are set as $\Omega_{R}(t)=A\sin(\omega t-\pi/2)$, $\Delta(t)=A\cos(\omega t-\pi/2)$, and $\phi=\phi_{2}$, which corresponds to the mixing angle $\theta(t)=\omega t-\pi/2$. During this stage, the mixing angle $\theta(t)$ varies from $0$ to $\pi$. By using Eq.~(\ref{eq6}), the unitary operators $V(\tau/2,\tau/2)$ and $V(3\tau/2,\tau/2)$ can be obtained as
\begin{align}
\begin{split}
&V\left(\frac{\tau}{2},\frac{\tau}{2}\right)=\left(
       \begin{array}{ccc}
         1 &0\\
         0 &1\\
       \end{array}
     \right),\\
&V\left(\frac{3\tau}{2},\frac{\tau}{2}\right)=\left(
       \begin{array}{ccc}
         0 &-e^{i(\frac{A\tau}{2}-\phi_{2})}\\
         e^{i(-\frac{A\tau}{2}+\phi_{2})} &0\\
       \end{array}
     \right).
\end{split}
     \label{eqa7}
\end{align}
The approximate propagator of the second stage in the rotating frame is given by Eq.~(\ref{eq17}). By using Eq.~(\ref{eq5}), the propagator of the second stage in the basis $\{\ket{0},\ket{1}\}$ can be obtained as
\begin{align}
\begin{split}
U\left(\frac{3\tau}{2},\frac{\tau}{2}\right)&=V\left(\frac{3\tau}{2},\frac{\tau}{2}\right)U_{\text{rot}}^{m}\left(\frac{3\tau}{2},\frac{\tau}{2}\right)V^{\dag}\left(\frac{\tau}{2},\frac{\tau}{2}\right)\\
&=\left(
       \begin{array}{ccc}
         0 & -e^{i[\frac{A\tau}{2}-\phi_{2}+\beta(3\tau/2,\tau/2)]}\\
        e^{i[-\frac{A\tau}{2}+\phi_{2}-\beta(3\tau/2,\tau/2)]} & 0\\
       \end{array}
     \right),
\end{split}
     \label{eqa8}
\end{align}
in which $\beta(3\tau/2,\tau/2)=\int^{3\tau/2}_{\tau/2}M(t)\sin(\omega t)dt$.

In the third stage $t\in{[3\tau/2,T]}$, the Rabi frequency, detuning, and phase of the control field are set as $\Omega_{R}(t)=A\sin(\omega t+\pi/2)$, $\Delta(t)=A\cos(\omega t+\pi/2)$, and $\phi=\phi_{3}$, which corresponds to the mixing angle $\theta(t)=\omega t+\pi/2$. Up to an irrelevant global phase, this stage can be equivalently regarded as having the mixing angle varying from $0$ to $\pi/2$. By using Eq.~(\ref{eq6}), the unitary operators $V(3\tau/2,3\tau/2)$ and $V(T,3\tau/2)$ can be obtained as
\begin{align}
\begin{split}
&V\left(\frac{3\tau}{2},\frac{3\tau}{2}\right)=\left(
       \begin{array}{ccc}
         1 &0\\
         0 &1\\
       \end{array}
     \right),\\
&V\left(T,\frac{3\tau}{2}\right)=\left(
       \begin{array}{ccc}
         \frac{\sqrt{2}}{2}e^{-i\frac{A\tau}{4}} &-\frac{\sqrt{2}}{2}e^{i(\frac{A\tau}{4}-\phi_{3})}\\
         \frac{\sqrt{2}}{2}e^{i(-\frac{A\tau}{4}+\phi_{3})} &\frac{\sqrt{2}}{2}e^{i\frac{A\tau}{4}}\\
       \end{array}
     \right).
\end{split}
     \label{eqa9}
\end{align}
The approximate propagator of the third stage in the rotating frame is given by Eq.~(\ref{eq19}). By using Eq.~(\ref{eq5}), the propagator of the third stage in the basis $\{\ket{0},\ket{1}\}$ can be obtained as
\begin{align}
\begin{split}
U\left(T,\frac{3\tau}{2}\right)&=V\left(T,\frac{3\tau}{2}\right)U_{\text{rot}}^{m}\left(T,\frac{3\tau}{2}\right)V^{\dag}\left(\frac{3\tau}{2},\frac{3\tau}{2}\right)\\
&=\left(
       \begin{array}{ccc}
         \frac{\sqrt{2}}{2}e^{i[-\frac{A\tau}{4}+\beta(T,3\tau/2)]} &-\frac{\sqrt{2}}{2}e^{i[\frac{A\tau}{4}-\phi_{3}-\beta(T,3\tau/2)]}\\
         \frac{\sqrt{2}}{2}e^{i[-\frac{A\tau}{4}+\phi_{3}+\beta(T,3\tau/2)]} &\frac{\sqrt{2}}{2}e^{i[\frac{A\tau}{4}-\beta(T,3\tau/2)]}\\
       \end{array}
     \right),
\end{split}
     \label{eqa10}
\end{align}
in which $\beta(T,3\tau/2)=\int^{T}_{3\tau/2}M(t)\sin(\omega t)dt$.

By substituting Eqs.~(\ref{eqa6}), (\ref{eqa8}) and (\ref{eqa10}) into the relation $U(T,0)=U(T,3\tau/2)U(3\tau/2,\tau/2)U\left(\tau/2,0\right)$, we obtain the propagator describing the entire measurement procedure for the quadrature component, as given in Eq.~(\ref{eq21}).

\end{document}